\documentclass[a4paper,11pt]{article}

\usepackage{jheppub} 

\usepackage[T1]{fontenc} 

\newcommand{\m}{m_1}
\newcommand{\n}{m_2}

\newcommand{\beq}{\begin{equation}}
\newcommand{\eeq}{\end{equation}}
\newcommand{\eea}{\end{eqnarray}}
\newcommand{\ba}{\begin{array}}
\newcommand{\ea}{\end{array}}
\newcommand{\bit}{\begin{itemize}}
\newcommand{\eit}{\end{itemize}}

\newcommand{\complesso}{{\ \hbox{{\rm I}\kern-.6em\hbox{\bf C}}}}
\newcommand{\reale}{{\hbox{{\rm I}\kern-.2em\hbox{\rm R}}}}

\newcommand{\1}{ \,  \raisebox{+0.14em}{{\hbox{{\rm \scriptsize ]}} \raisebox{-0.2em}{\kern-.8em\hbox{1}}}} \, }
\newcommand{\p}{\partial}

\newcommand{\eal}[1]{\begin{equation} \begin{aligned} #1 \end{aligned}\end{equation}}

\title{\boldmath Meronic Einstein-Yang-Mills black hole in 5D and gravitational spin from isospin effect}

\author[a]{Fabrizio Canfora,}
\author[b,a]{Andr\'es Gomberoff,}
\author[c]{Seung Hun Oh,}
\author[b]{Francisco Rojas,}
\author[d]{Patricio Salgado-Rebolledo}

\affiliation[a]{Centro de Estudios Cient\'{\i}ficos (CECS), Casilla 1469, Valdivia, Chile}
\affiliation[b]{Facultad de Ingenier\'ia y Ciencias, UAI Physics Center, Universidad Adolfo Ib\'a\~nez,\\ Avda. Diagonal Las Torres 2640, Pe\~nalol\'en, Santiago, Chile}
\affiliation[c]{Seoul National University of Science and Technology, Seoul, 01811, Korea}
\affiliation[d]{Instituto de F\'isica, Pontificia Universidad Cat\'olica de Valpara\'iso, Casilla 4059, Valpara\'iso, Chile}

\emailAdd{canfora@cecs.cl}
\emailAdd{andres.gomberoff@uai.cl}
\emailAdd{shoh.physics@gmail.com}
\emailAdd{francisco.rojasf@uai.cl}
\emailAdd{patricio.salgado@pucv.cl}

\abstract{We construct an analytic black hole solution in $SU(2)$ Einstein-Yang-Mills theory in five dimensions supporting a Meron field. The gauge field is proportional to a
pure gauge and has a non-trivial topological charge. The would-be singularity at the Meron core gets shielded from the exterior by the black hole horizon. The metric has only one integration constant, namely, its ADM mass, which is shown to be finite once an appropriate boundary term is added to the action. The thermodynamics is also worked out, and a first-order phase transition, similar to the one occurring in the Reissner-Nordstr{\"o}m case is identified. We also show that the solution produces a \mbox{\textit{spin from isospin effect}}, \emph{i.e.}, even though the theory is constructed out of bosons only, the combined system of a scalar field and this background may become fermionic. More specifically, we study scalar excitations in this purely bosonic background and find that the system describes fermionic degrees of freedom at spatial infinity. Finally, for the asymptotically $AdS_{5}$ case, we study its consequences in the context of the AdS/CFT correspondence.}

\begin{document}
\maketitle
\flushbottom

\section{Introduction}

It has long been recognized that topological solitons are key ingredients of the non-perturba\-tive sector of quantum field theories (for reviews see \cite{manton2004topological, greensite2010introduction}). In some cases, the asymptotic form of these solitons demand that a spatial rotation must be compensated by  a Yang-Mills gauge transformation. This remarkable phenomenon, which usually goes under the name of \textit{Goldhaber-Jackiw-Rebbi-Hasenfratz-'t\ Hooft effect}
\cite{Jackiw:1976xx, Hasenfratz:1976gr, Goldhaber:1976dp} (also
known as \textit{spin from isospin effect}),  implies that a
composite of a bosonic field and solitons of this type, can become fermionic. That is, starting from a purely bosonic theory, fermionic excitations arise nonetheless. In
situations in which the gravitational back-reaction is not negligible, however, this property of topologically non-trivial solitons has not been thoroughly studied (see, for instance, \cite{Canfora:2012ap}). In this article we report the existence of a new spherically symmetric black hole solution for the Einstein-Yang-Mills
system\footnote{For related works on Einstein-Yang-Mills black holes see \cite{Cordero:1976jc,Bizon:1990sr,kunzle1990spherically,Volkov:1998cc,Brihaye:2002jg,Okuyama:2002mh,Gibbons:2006wd,Brihaye:2007jua,Winstanley:2008ac,Mazharimousavi:2008ap,Bostani:2009zf,Agop:2010zz,Winstanley:2015loa,Volkov:2016ehx,Shepherd:2016ily,Sadeghi:2018ylh} and references therein.} in which the gauge field is  a Meronic configuration with such solitonic structure.
 
On flat spaces, Merons
(firstly introduced in \cite{deAlfaro:1976qet}) can be thought of as building blocks of instantons, and are relevant configurations when studying confinement \cite{Callan:1977gz,Callan:1977qs,Glimm:1977sx,Callan:1978bm,Actor:1979in,Lenz:2003jp}. However, they suffer from a problem: they are singular at their core. 
An interesting feature of the \emph{gravitating} Meron found here is that its center lies inside the black hole horizon, preventing its singular behavior to affect the exterior. 
Moreover, it also exhibits the \emph{spin from isospin effect}.

To obtain the meronic black hole we study analytic, Meron--type solutions of the Einstein-Yang-Mills field equations, using the hedgehog Ansatz introduced in \cite{Canfora:2013mrh, Canfora:2013osa, Canfora:2013xja,Chen:2013qha,Canfora:2014aia,Canfora:2014jka,Canfora:2014jsa,Canfora:2015xra,Chen:2015bek,Ayon-Beato:2015eca,Canfora:2015hxa,Canfora:2016spb,Tallarita:2017bks,Canfora:2017ivv,Canfora:2017gno,Canfora:2017yio}. The Yang-Mills $SU(2)$ gauge potential is $A=\lambda \widetilde{A}$, where $\widetilde{A}$ is pure gauge. Due to the fact that in abelian gauge theories these types of potentials are necessarily trivial, the existence of Merons is a genuine
non-abelian feature. Indeed, for $\lambda\neq0,1$, the Meron configuration
has a non-zero field strength: $F=\lambda(\lambda
-1)[\widetilde{A},\widetilde{A}]$. 

In the context of the AdS/CFT correspondence, one studies strongly coupled phenomena using well controlled
methods of weakly coupled gravitational theories. Thus, whenever we have a new
gravitational tool at our disposal, it is natural to ask what can it do for the dynamics of strongly interacting matter. Having in mind the \textit{gravitational spin from isospin} effect we describe here, we envisage the possibility of using it as a way to describe fermionic observables in the CFT using only bosonic fields in the bulk.

As a first example, we study the equation of motion for a scalar field isospin doublet in the Meron black hole background. We show how its asymptotic behavior, near the AdS boundary, can be recovered from the equation of motion of a free massive two-spinor on the boundary geometry. 

As a second, and stronger argument in favor of the holographic applicability of this new solution, we study the Fock space of free particle states in the bulk geometry and compare it with that of the boundary CFT operators, following the approach of \cite{Witten:1998qj,Balasubramanian:1998de,Balasubramanian:1998sn,Balasubramanian:1999ri}. We show that, when expanding the CFT operators in terms of spherical harmonics at the $AdS_5$ boundary (the $\mathbb{R}\times S^3$), the angular momentum quantum numbers acquire half-integer values in the presence of the Meron black-hole. Thus, necessarily implying a fermionic character of such operators when the Meron is turned on.

Also within AdS/CFT, strongly coupled systems on $\mathbb{R}^{n}$ have been the ones most
intensively studied; black holes in this case are black branes whose horizons share the topology of $\mathbb{R}^{n}$ with the boundary theory. The study of
black holes with spherical horizons have received, however, considerably less
attention. For example, although black branes in AdS have been extensively used in the study of holographic superconductors (for a recent review see
\cite{Hartnoll:2016apf}), spherical black holes also exhibit the same kind of instability required to describe superconductors \cite{Gubser:2008px}. The new black hole solution found here thus
enlarges the set of systems with spherical symmetry that can be addressed
using holography. Moreover, since the solution we present also exhibits a first-order phase transition, the system could hint at novel holographic features. Another amusing feature of our black hole solution is that it exhibits a logarithmic dependence in the radial coordinate (see Eq. \eqref{eq:f-gen}). Since the radial coordinate of $AdS_5$ is dual to the RG scale of the boundary theory, our solution hints a logarithmic running of the coupling, similar to that found by Klebanov and Strassler \cite{Klebanov:2000hb}, and originally pointed out in \cite{Klebanov:1999rd}.

The paper is organized as follows. In section \ref{sec:meron} we study $SU(2)$ gauge fields of Meron-type on spherically symmetric spacetimes and solve the Yang-Mills equations. Using the generalized hedgehog ansatz,  we find the explicit form of the gauge field and show that it cannot be continuously ``turned off". In section \ref{sec:bh} we turn our attention to the gravitational part of the system and present a black hole solution of the Einstein equations that supports the Meron gauge field. We show that the ADM mass can be consistently computed by noting that, due to the presence of the Meron, the background solution is not Minkowski space, but the zero mass solution of the black hole spectrum. In section \ref{sec:sfi} we show how a gravitating Meron induces a spin from
isospin effect, implying that a scalar particle moving in the meronic black hole background has half-integer angular momentum eigenvalues, and finalize with envisaging the possibility of using this effect in the context of the
AdS/CFT correspondence.

\section{Gravitating Meron gauge field}
\label{sec:meron}

We consider the $SU(2)$ Einstein-Yang-Mills action in five dimensions with
cosmological constant
\begin{equation}
I=\frac{1}{16\pi}\int d^{5}x\sqrt{-g}\left(  \frac{R-2\Lambda}{G}+\frac
{1}{e^{2}}\mathrm{Tr}\left[  F^{\mu\nu}F_{\mu\nu}\right]  \right)  ,
\label{skyoo}%
\end{equation}
where $R$ is the Ricci scalar, $F_{\mu\nu}=\partial_{\mu}A_{\nu}-\partial_{\nu
}A_{\mu}+\left[  A_{\mu},A_{\nu}\right]  $ is the field strength of the gauge
field $A_{\mu}$, $G$ is Newton's constant, $\Lambda$ is the cosmological
constant and $e$ is the Yang-Mills coupling. In our conventions $c=\hbar=1$.
Einstein equations are given by
\begin{equation}
G_{\mu\nu}+\Lambda g_{\mu\nu}=8\pi G\;T_{\mu\nu} \ , \label{einstein}%
\end{equation}
where $G_{\mu\nu}=R_{\mu\nu}-\frac{1}{2}g_{\mu\nu}R$ is the Einstein tensor, and $T_{\mu\nu}$ is the stress-energy tensor of the Yang-Mills field:

\begin{equation}
T_{\mu\nu}=\frac{1}{4\pi e^{2}}\mathrm{Tr}\left(  -F_{\mu\alpha}F_{\nu\beta
}g^{\alpha\beta}+\frac{1}{4}g_{\mu\nu}F^{\alpha\beta}F_{\alpha\beta}\right)
. \label{T1}%
\end{equation}
The Yang-Mills equations, on the other hand, read
\begin{equation}
\nabla_{\nu}F^{\mu\nu}+\left[  A_{\nu},F^{\mu\nu}\right]  =0\ , \label{YM1}%
\end{equation}
where $\nabla^{\mu}$ is the Levi-Civita covariant derivative.

The gauge field $A_{\mu}=-iA_{\mu}^{a}t_{a}$ and therefore the field-strengh $F_{\mu\nu}=-iF^{a}_{\mu\nu}t_a$ take values on the $\mathfrak{su}%
(2)$ algebra, where the $t_{a}$ ($a=1,2,3$) are Hermitian generators
satisfying
\[
\lbrack t_{a},t_{b}]=i\epsilon_{\;ab}^{c}t_{c}\quad,\quad\mathrm{Tr}[t_{a}%
t_{b}]=\frac{1}{2}\delta_{ab}\,.
\]
The action is invariant under gauge transformations of the form
\begin{equation}
\label{gtransformation}A_{\mu}^{\prime}=U A_{\mu} U^{-1}-\partial_{\mu}\,UU^{-1}
\, ,\quad U=e^{-i\alpha^{a} \,t_{a}} \in SU(2)\,.
\end{equation}

\subsection{Ansatz}
\label{sec:ansatz}

We will consider the usual spherically symmetric ansatz for the five-dimensional space-time metric:

\begin{equation}
ds^{2}=-f^{2}(r)dt^{2}+\frac{1}{f^{2}(r)}dr^{2}+\frac{r^{2}}{4}d\Omega_{S^{3}%
}^{2}\,, \label{eq:Metrica1}%
\end{equation}
where $t$ stands for the time coordinate, $r$ corresponds to the radial coordinate, and $d\Omega_{S^3}^2$ denotes the line element of the unit three-sphere. The group $SU(2)$ is diffeomorphic to $S^{3}$. Therefore $d\Omega^{2}_{S^{3}}$ can be written as (see Appendix \ref{sec:su2})
\begin{equation}
\label{metrics3}d\Omega^{2}_{S^{3}}=-\frac{1}{2}\mathrm{Tr}[dUU^{-1}\otimes
dUU^{-1}]\,,
\end{equation}
where $dUU^{-1}=-i \omega_{R}^{a} t_{a}$ corresponds to the right-invariant
Maurer-Cartan form on $SU(2)$, and whose components satisfy
\begin{equation}
\label{mceqR}d \omega_{R}^{a}=\frac{1}{2}\epsilon^{a}_{\;bc}\omega_{R}^{b}\wedge\omega_{R}^{c}\,.
\end{equation}

For the gauge field we will consider Meron-like configurations, \emph{i.e.},
connection one-forms $A=A_{\mu}dx^{\mu}$ that are proportional to a pure gauge
field \cite{deAlfaro:1976qet}. According to \eqref{gtransformation} this means
\begin{equation}
A=-\lambda dUU^{-1}\quad,\quad   \lambda \neq 0,1, \label{merona}%
\end{equation}
which is equivalent to
\begin{equation}
A=i\lambda\omega_{R}^{a}t_{a}\,. \label{merona1}%
\end{equation}
The corresponding curvature two-form is given by $F=dA+A\wedge A=\frac{1}{2}F_{\mu\nu}dx^\mu dx^\nu$. For a
gauge field of the form \eqref{merona1} it takes the simple form
\begin{equation}
F=\frac{\lambda-1}{2\lambda}\left[  A,A\right]  \,. \label{eq:F}%
\end{equation}

\subsection{Yang-Mills Equations}
\label{sec:ym}

In the case of a Meron gauge field, Yang-Mills equations (\ref{YM1})  can be easily solved when written in terms of forms:
\begin{equation}
d*F+[A,*F]  = 0\,, \label{YM2-1}%
\end{equation}
where $*F$ denotes the Hodge dual of $F$. In the following, it will prove convenient to define a set of f\"unfbeins
$e^{A}$, $A=0,\ldots,4$, such that $ds^{2}=\eta_{AB}e^{A}e^{B}$ with
$\eta_{AB}=\mathrm{diag}\left(  -,+,+,+,+\right)  $. The natural choice is
\begin{align}
e^{0}  &  = f(r)dt\,,\nonumber\\
e^{1}  &  = \frac{1}{f(r)}dr\,,\label{eq:vielbeins}\\
e^{a+1}  &  = \frac{r}{2}\omega_{R}^{a}\,.\nonumber
\end{align}
The gauge connection (\ref{merona1}) can be written in terms of the
angular f\"unfbeins as
\begin{equation}
A=\frac{2i\lambda}{r}e^{a+1}t_{a}\,.
\end{equation}
Therefore, in the f\"unfbein basis $A=A_{A}e^{A}$, the components of the Meron gauge field take the
simple form
\begin{equation}
A_{A}=\frac{2i\lambda}{r}\delta_{A}^{a+1}t_{a}\,.
\end{equation}
Similarly, the curvature two-form (\ref{eq:F}) can be
written in terms of the f\"unfbein basis as $F=\frac{1}{2}F_{AB}e^{A}\wedge
e^{B}$, with components
\begin{equation}
F_{AB}=-\frac{4i\lambda\left(  \lambda-1\right)  }{r^{2}}\epsilon_{\;bc}%
^{a}\delta_{A}^{b+1}\delta_{B}^{c+1}t_{a}\,.
\end{equation}
The Hodge dual of $F$, on the other hand, takes the form
\begin{equation}
\label{hodgeF}*F=-\frac{4i\lambda\left(  \lambda-1\right)  }{r^{2}}\,t_{a}\,
e^{0}\wedge e^{1}\wedge e^{a+1}\,.
\end{equation}
Using \eqref{eq:vielbeins} and \eqref{hodgeF}, equation (\ref{YM2-1}) reduces
to
\begin{equation}
\frac{2i\lambda\left(  \lambda-1\right)  }{r}\left(  d\omega_{R}^{a}%
-\lambda\epsilon_{\;bc}^{a}\omega_{R}^{b}\wedge\omega_{R}^{c}\right)
\,t_{a}\;e^{0} \wedge e^{1}=0\,,
\end{equation}
which, due to the Maurer Cartan equation \eqref{mceqR}, is solved for
\begin{equation}
\lambda=\frac{1}{2}\ . \label{eq:solym}%
\end{equation}

\subsection{Generalized hedgehog ansatz}
\label{sec:gen}

The group element $U\in SU(2)$ that corresponds to the Meron gauge
potential \eqref{merona} can be explicitly written using the generalized hedgehog ansatz \cite{Canfora:2013mrh}. It is interesting to note that such an
approach (subsequently worked out in \cite{Canfora:2013osa, Canfora:2013xja,Canfora:2014aia,Canfora:2014jka, Chen:2013qha,Canfora:2015xra,Chen:2015bek,Ayon-Beato:2015eca,Canfora:2014jsa,Canfora:2015hxa,Canfora:2016spb,Tallarita:2017bks,Canfora:2017ivv,Canfora:2017gno,Canfora:2017yio}) was
originally developed to analyze the Skyrme and Einstein-Skyrme models, but it also works in the case of Yang-Mills and Einstein-Yang-Mills theories with almost
no modifications (as one can see comparing \cite{Canfora:2013osa}, \cite{Ayon-Beato:2015eca} with
\cite{Canfora:2012ap}, \cite{Canfora:2017yio}). This approach is specially designed for
situations in which the group element cannot be deformed continuously to
the identity. 

In order to write down the ansatz for $U$, we adopt the standard parametrization for $SU(2)$ elements:
\begin{equation}
U(x^{\mu})=e^{-i\alpha^{a}t_{a}}=\mathrm{cos}\frac{\alpha}{2}(x^{\mu
})\mathbb{\mathbf{I}}-2in^{a} t_{a}\,\mathrm{sin}\frac{\alpha}{2}(x^{\mu
})\,, \label{standnorm-1}%
\end{equation}
where $\mathbb{\mathbf{I}}$ is the $2\times2$ identity and $n^{a}$ is a unit
vector that can be written in the form
\begin{equation}
n^{1}=\mathrm{sin}\Theta\mathrm{cos}\Phi\,\,\,,\,\,\,n^{2}=\mathrm{sin}%
\Theta\mathrm{sin}\Phi\,\,\,,\,\,\,n^{3}=\mathrm{cos}\Theta\,\,.
\label{standard2}%
\end{equation}
The generalized hedgehog ansatz corresponds to the following choice for the
functions $\alpha$, $\Theta$, $\Phi$ in terms of Euler variables:
\begin{equation}
\Phi=\frac{\varphi-\psi}{2}\ ,\ \ \tan\Theta= \tan \dfrac{\theta}{2}  \csc \dfrac{\varphi+\psi}{2}  \, ,\ \ \cot
\frac{\alpha}{2}=\sec\Theta \tan \frac{\varphi+\psi}{2}  \ . \label{pions2.25-1}%
\end{equation}
This leads to the following form for the components of the right-invariant
Maurer-Cartan form on SU(2):
\begin{align}
\omega_{R}^{1}  &  =-\mathrm{sin}\psi d\theta+\mathrm{cos}\psi\,
\mathrm{sin}\theta d\varphi\,,\nonumber\\
\omega_{R}^{2}  &  =\mathrm{cos}\psi d\theta+\mathrm{sin}\psi\,
\mathrm{sin}\theta d\varphi\,,\label{lmc}\\
\omega_{R}^{3}  &  =d\psi+\mathrm{cos}\theta d\varphi\,,\nonumber
\end{align}%
\begin{equation}
\label{eq:angles}
0\leq\psi<4\pi\,,\,\,0\leq\theta<\pi\,,\,\,0\leq\varphi<2\pi\,,
\end{equation}
which clearly satisfy \eqref{mceqR}. Using \eqref{lmc}, and defining the coordinates $x^{m}=\left\{ \psi,\theta,\phi\right\}$, the metric for the three-sphere \eqref{metrics3} takes the form
\begin{equation}
d\Omega_{S^{3}}^{2}=h_{mn}dx^{m}dx^{n}\;,\quad h_{mn}=\frac{1}{4}\left(\begin{array}{ccc}
1 & 0 & \cos\theta\\
0 & 1\\
\cos\theta & 0 & 1
\end{array}\right) \, . \label{metrics3-2}%
\end{equation}
while the components of the gauge field \eqref{merona1} read
\begin{equation}\label{eq:Ameron}
\begin{array}{lcl}
&&A_{t}  =A_{r}=0\,,\\[5pt]
&&A_{\psi}  =\dfrac{i}{2}\,t_{3}\,,\\[5pt]
&&A_{\theta}  =\dfrac{i}{2}\left(  -\sin\psi\,t_{1}+\cos\psi\,t_{2}\right)
\,,\\[5pt]
&&A_{\varphi}  =\dfrac{i}{2}\left( \cos\psi \sin\theta\,t_{1}+\sin\psi \sin\theta
\,t_{2}+\cos\theta\,t_{3}\right)\,,
\end{array}
\end{equation}
where we have also used \eqref{eq:solym},  \emph{i.e.}, $\lambda=1/2$.

It is worth emphasizing that the group valued element in Eqs.
(\ref{standnorm-1}), (\ref{standard2}) and (\ref{pions2.25-1}) (which is the
basic building block of the Meron ansatz) is topologically non-trivial as
it has a non-vanishing winding number along the $r=\mathrm{const}$ hypersurfaces
of the metric \eqref{eq:Metrica1}. In fact, the second Chern number associated with the Meron gauge field \eqref{eq:Ameron} is given by
\begin{equation}\label{chern}
C_2= \frac{1}{8\pi^{2}}\int_{S^{3}}\mathrm{Tr}\left[  A\wedge dA+\frac{2}{3}A \wedge A \wedge A \right]=\frac{1}{48\pi^{2}}\int_{S^{3}}\mathrm{Tr}\left[  \left(dU U^{-1}\right)^3 \right]=\frac{1}{2}\, .
\end{equation}
Hence, the group valued element defined by Eqs. (\ref{standnorm-1}), (\ref{standard2})
and (\ref{pions2.25-1}) cannot be continuously deformed to the identity.\footnote{The integral \eqref{chern} is the Pontryagin index. For Yang-Mills instantons, it defines a topological charge that takes integer values. Merons, on the other hand, are solutions of Yang-Mills equations with one-half unit of topological charge \cite{Actor:1979in}.}

\section{Black hole solution}
\label{sec:bh}

As we have expressed the metric on $S^3$ in the form \eqref{metrics3-2}, the space-time metric \eqref{eq:Metrica1} takes the form
\begin{equation}  \label{eq:Metrica2}
ds^{2}=-f^{2}(r)dt^{2}+\frac{1}{f^{2}(r)}dr^{2}+\frac{r^{2}}{4}\left(
d\psi^{2}+2\cos\theta d\psi d\varphi+d\theta^{2}+d\varphi^{2}\right)  .
\end{equation}
Using this and \eqref{eq:Ameron}, it is straightforward to check that Einstein equations \eqref{einstein}
admit an exact solution with
\begin{equation}
\label{eq:f-gen}f^{2}(r)=1-\frac{8MG}{3\pi r^{2}}-\frac{1}{r^{2}%
}\left[ \left(1-\frac{\Lambda}{6}\ell^{2}\right)\ell^{2}+ r_0^2\mathrm{Log}\frac{r^{2}}{\ell^{2}}\right]  -\frac{\Lambda}%
{6}r^{2}\,,
\end{equation}
where $M$ is an integration constant that will later be  identified with the mass of the black hole, and
\begin{equation}
	\ell^{2}=\frac{3}{2\Lambda}\left(  1 - \sqrt{1-\frac{4\Lambda r^2 _0 }{3}%
}\right), \ \ \ r_0^2=\frac{G}{2e^2} \, .  \label{l2}
\end{equation}
 The solution exists provided
$$
\frac{2\Lambda G}{3e^2}\leq 1 \, .
$$
This condition is met trivially for $\Lambda\leq 0$, whereas for $\Lambda > 0$ it implies that the solution exists only if the coupling of the Yang-Mills field is sufficiently large.
The solution is a black hole, whose horizon is determined by the equation $f^2(r_+)=0$. 
Note that  $\ell$ in Eq.\eqref{l2} has a smooth limit  $\ell\rightarrow r_0$ as the cosmological constant vanishes. The parameter $M$ has been designed so that $M=0$ is the smallest possible value for the solution not to have naked singularities. For $M=0$ the horizon is located at its minimum value, which is $r_+=\ell$, and the black hole  is extreme, as $(f^{2})^{\prime}|_{r_+}=0$.

\subsection{Mass}
\label{sec:mass}

As previously announced, the mass of the black hole solution is precisely the parameter $M$ in \eqref{eq:f-gen}. Unlike what happens, for instance, with the
Reissner-Nordstr{\"o}m black hole, this is the only parameter present in the solution. There is no integration constant associated with the gauge potential, and therefore there is no way to ``turn off" the Yang-Mills field: the
parameter $\lambda$ in front of the gauge potential \eqref{merona} is uniquely
fixed\footnote{If $\lambda$ was not fixed, it would play a
similar role to the electric charge $Q$ appearing as a free parameter in the
ansatz for the electric gauge potential in the electric Reissner-Nordstr{\"o}m
black hole. In the usual case, the gauge field solves the Maxwell equations for
any value of $Q$. Thus, in the Reisssner-Nordstr{\"o}m black hole one can
continuously turn off $Q$.} by the Yang-Mills field equations to be
$\lambda=1/2$.  Therefore, one expects that the vacuum solution of zero energy needed to
properly define the ADM mass of this black hole should not coincide with the
naive vacuum solution in which the gauge potential is vanishing. The following
computations confirm this point of view very clearly.

In the asymptotically flat case ($\Lambda=0$), the ADM mass can be defined, and is given by \cite{Arnowitt:1962hi} (see \cite{Hanson:1976cn} for a detailed analysis)
\begin{equation}\label{ADM}
M^{ADM}={\rm lim}_{r\rightarrow\infty}\;\frac{1}{16\pi G}\int_{S^3}
\sqrt{\gamma}dS \;\hat{n}^{i}\gamma^{jk}\left(  \text{$\partial$%
}_{k}\gamma_{ij}-\partial_{i}\gamma_{jk}\right)\,,
\end{equation}
where $\hat{n}$ is the radial unit vector, $dS$ stands for surface element on $S^3$, $\gamma$ is the spatial part of \eqref{eq:Metrica2}, \emph{i.e.},
\begin{equation}\label{spatial}
\gamma_{ij}dx^{i}dx^{j}=\frac{1}{f^2 \left(  r\right) %
}dr^{2}+\frac{r^{2}}{4}\left(  d\text{$\psi$}^{2}+2\cos\theta d\psi
d\varphi+d\theta^{2}+d\varphi^{2}\right)  \ ,
\end{equation}
and $i,j=1,2,3,4$. This leads to the following divergent expression for \eqref{ADM}
\begin{equation}\label{ADM2}
M^{ADM}=\mathrm{lim}_{\;r\rightarrow\infty}\;\frac{3\pi r^{2}}{8G}\left(
\frac{1}{f^2 (r)}-1\right) \,.
\end{equation}
Even though this quantity is infinite, the divergent term does not depend on $M$. This means that the mass can be made finite by defining the
$M=0$ black hole as the vacuum configuration. To do so one defines a new action principle by subtracting the action evaluated in the vacuum solution to the original one.  
The resulting mass is
\[
M^{ADM}-M_{\mathrm{vacuum}}^{ADM}:=\;M\,.
\]
An analog argument, along the lines of \cite{Henneaux:1984xu,Gomberoff:2003ea}, for example, may be carried out when the cosmological constant does not vanish. We will not go through that analysis here, however, the consistency of this definition with the first law of thermodynamics (see next subsection)  it is  indeed reassuring.

Note that the $M=0$ solution is not to be interpreted as the  global vacuum of the theory. It has been defined as the lowest energy configuration  included  in the family of solutions discussed here. It is not characterized by a symmetry enhancement of any sort. A similar situation is encountered, for example, in three dimensional gravity with negative cosmological constant. There, the BTZ black hole solution \cite{Banados:1992wn, Banados:1992gq} defined as having zero mass does not correspond to AdS space, which is only recovered when setting the mass parameter $M=-1$. The reason for choosing such a definition is that for ${-1<M<0}$ the resulting geometries contain a conical naked singularity. There is, however, an important difference between that case and the meronic black hole: in the latter there is no value of the mass parameter for which a maximally symmetric solution is recovered. This was to be expected due to the topological nature of this family of solutions. They have a non--vanishing Chern invariant \eqref{chern} and therefore cannot be  continuously deformed to a maximally symmetric vacuum with a vanishing Chern invariant. This non-perturbative property of the meronic solutions can also be seen from the fact that the limit to (A)dS or Minkowski are obtained in the strong coupling limit of $e\rightarrow\infty$.

Finally note, as we discussed above for the asymptotically flat case, that there is an infinite mass gap between the meronic black hole and the Minkowski vacuum. This should not be a cause of concern. It is precisely what happens, for instance, between the false vacuum and the true gravitational dS vacua described by  Coleman and De Luccia in \cite{Coleman:1980aw}. A phase transition to the true vacuum may only occur via bubble nucleation, which starts on a finite volume and then grows to cover the entire space in an infinite amount of time. One would expect that the meronic solutions could have been produced as primordial black holes, and that there should exist a mechanism (thermal or quantum mechanical) that will make them decay to the true, maximally symmetric, vacuum.

\subsection{Thermodynamics}\label{sec:thermo}

The thermodynamics of the meronic black hole is quite similar to the one of an electrically charged black hole with cosmological constant \cite{Louko:1996dw,Chamblin:1999tk}. As mentioned above, however,  in the meronic case there is no  integration constant analog to the electric charge that could be varied. There is, therefore, only one thermodynamical extensive parameter: the mass $M$. 

The temperature of the black hole may be computed most easily passing to the Euclidean metric by making the identification $t=i\tau$, so that
\begin{equation}\label{bhmetric}
ds^{2}=f^{2}(r)d\tau^{2}+\frac{1}{f^{2}(r)}dr^{2}+ r^2 d\Omega_{S^3}^2\,.
\end{equation}
As usual, the coordinate $\tau$ must have a well-defined periodicity $\beta$ for the metric to be regular. The temperature $T=\beta^{-1}$ is given by
\begin{equation}
T=\frac{(f^{2})^{\prime}|_{r_+}}{4\pi} \ ,
\end{equation}
Since the gravitational action we are using here is the Einstein-Hilbert one, the entropy is
\[
S=\frac{1}{4G}(\mbox{horizon's area})\,.
\]
We will analyze the case with $\Lambda=0$ first. In that case we can write  the mass, temperature, and the entropy of the black hole in terms of the horizon radius: $r_+$,
\begin{equation}
	M=\frac{3\pi}{8G}\left\{r_+^2-r_0^2\left(\log\left[\frac{r_+^2}{r_0^2}\right]+1\right)\right\}\,,
	\ \  T=\frac{r_+^2-r_0^2}{2\pi r_+^3}\,, \ \ \ S=\frac{\pi^2r_+^3}{2G}\,.
\end{equation}
As announced in the previous section, the definition of mass used here satisfies the first law of black hole thermodynamics,
$$
dM=TdS\,.
$$
Of course, the first law determines the mass up to a constant shift. We have adjusted that shift in \eqref{eq:f-gen} so that the solutions with $M<0$ have naked singularities. 

\begin{figure}[h] 
\begin{center}
\includegraphics[width=0.8\textwidth]{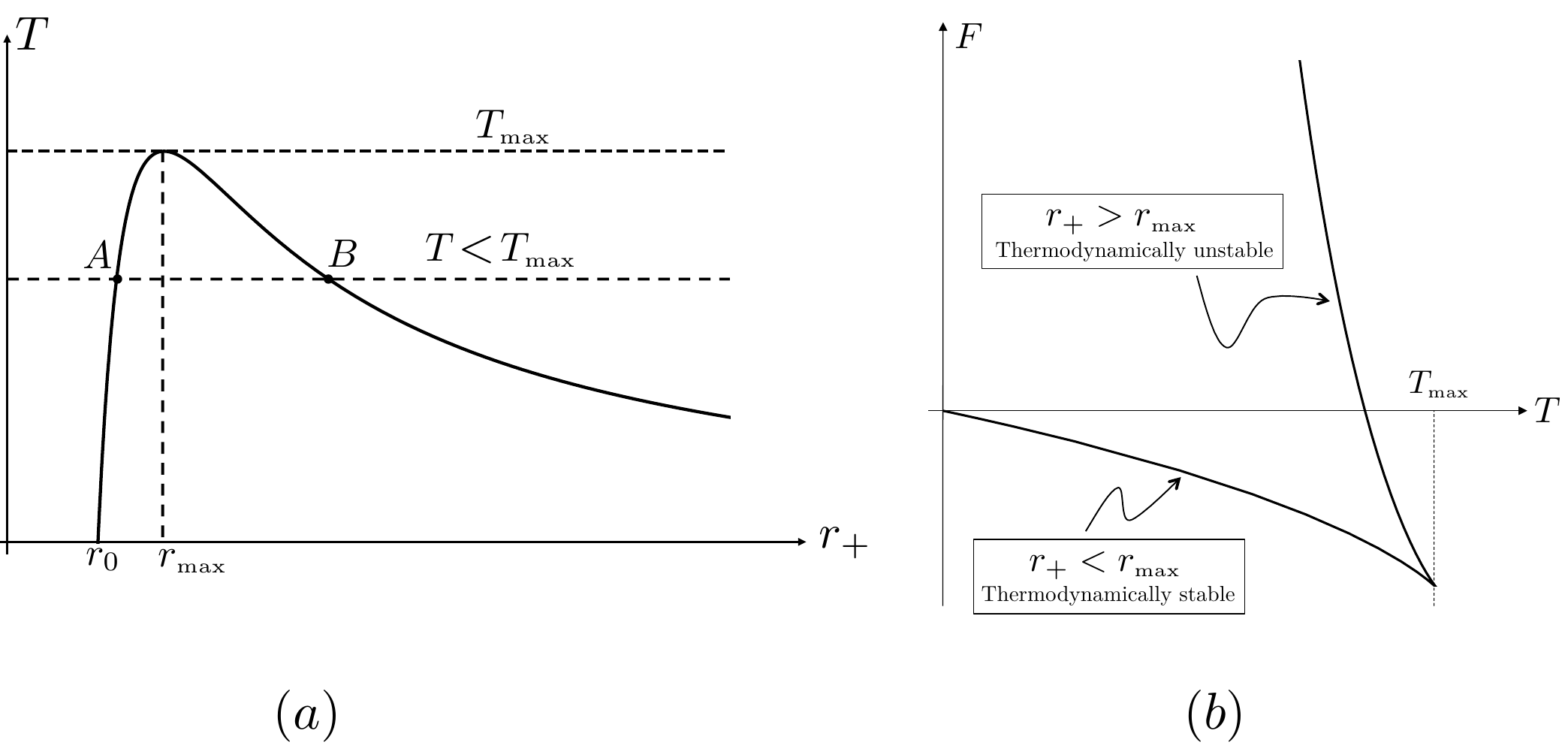}
\caption{Figure (a) is a graph of the temperature of the black hole solution as a function of the horizon radius $r_+$ in the case of vanishing cosmological constant $\Lambda=0$. The solution with $r_+=r_0$ is the smallest possible black hole, it corresponds to the vacuum solution $M=0$, and has $T=0$. For $T\in[0,T_{\mbox{\tiny max}}]$ there are two possible black holes, namely points $A$ and $B$ depicted in the figure, but only the one with smallest radius is thermodynamically stable. For $T>T_{\mbox{\tiny max}}$ there are no black hole solutions. Figure (b) is a graph of the Helmholtz free energy $F$ as a function of $T$, clearly showing the unstable region versus the stable one.}
\label{tr}
\end{center}
\end{figure}

The thermodynamical properties of the black hole solution in this case can be extracted from figure \ref{tr} (a), where a graph of temperature versus $r_+$ is shown. We see that, if the temperature is larger than
\[
T_{\mbox{\tiny max}}=\frac{1}{3\sqrt{3}\pi r_0}\, ,
\]
there is no black hole solution.  If $T=T_{\mbox{\tiny max}}$, a unique black hole of radius $r_{\mbox{\tiny max}}=\sqrt{3}r_0$ is formed. If we lower the temperature to the region $T\in[0,T_{\mbox{\tiny max}}]$, there are two possible black hole solutions. 
One with $r_+>r_{\mbox{\tiny max}}$, and one with $r_+<r_{\mbox{\tiny max}}$. Note, however, 
that black holes of the first kind are thermodynamically unstable, because they have a negative heat capacity. This can be seen from the graph. Indeed, as the entropy is a monotonically 
increasing function of $r_+$, it is clear that $\partial T/\partial S$ is positive for $r<r_{\mbox{\tiny max}}$ and negative for $r>r_{\mbox{\tiny max}}$. 
Therefore, black holes with $r_+=r_{\mbox{\tiny max}}$ are the biggest ones that are thermodynamically stable. These are also the more massive, because $M$ is also a monotonically increasing function of $r_+$ for $r_+>r_0$.  If a black hole with $r_+>r_{\mbox{\tiny max}}$ is formed in a reservoir at temperature $T$, say, at the point $B$ of Fig. \ref{tr} (a), it will evaporate until it reaches point $A$. There, it will stay in thermodynamical equilibrium with the reservoir. Finally, for $T=0$ we have the vacuum solution, with $r_+=r_0$.
Figure \ref{tr} (b) shows the Helmholtz free energy, $F=MT-S$ as a function of $T$. It clearly shows that the stable region is the one with lower $F$, corresponding to a horizon radius of $r_+<r_{\mbox{\tiny max}}$.

\begin{figure}[h] 
\begin{center}
	\includegraphics[width=0.8\textwidth]{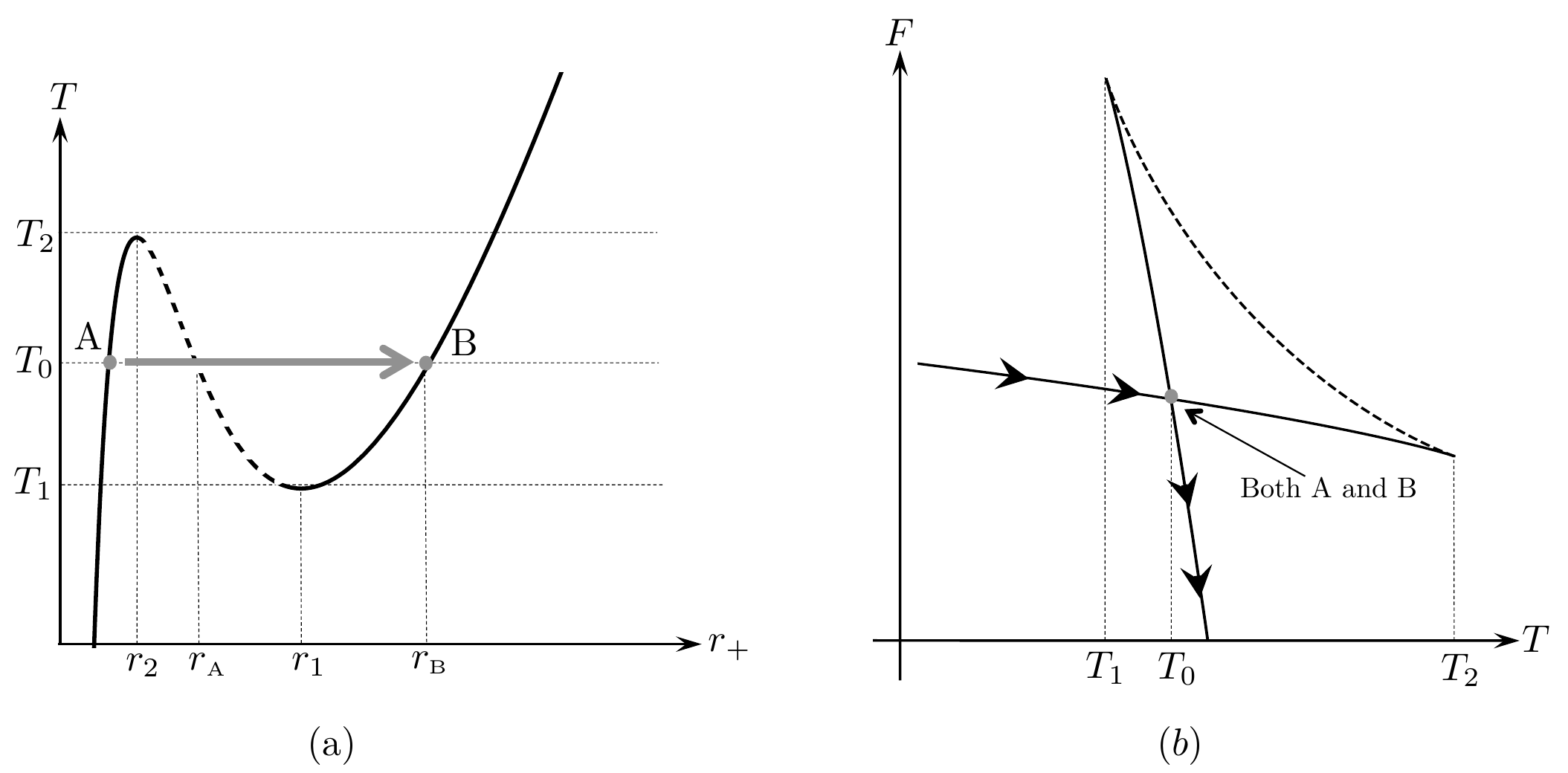}
\caption{Figure (a) is a graph of the temperature of the black hole solution as a function of the horizon radius $r_+$ in the case of negative cosmological constant with $|\Lambda| <\frac{1}{4r_0^2}$. Again, the solution with $T=0$ is the smallest possible black hole, and it corresponds to the vacuum solution $M=0$. As one increases the temperature, when $T=T_0$, there is a first-order phase transition from  $A$ to $B$. Figure (b) shows the three branches of the graph of Helmholtz free energy $F$ as a function of $T$. The dashed lines show thermodynamically unstable branches. As one increases the temperature, the system goes along the path of minimal Helmholtz free energy (arrows in figure (b)).  The free energy is continuous at $T_0$, but its derivative, the entropy, is not, showing that this is a first-order transition.}
\label{tr2}
\end{center}
\end{figure}

When the cosmological constant is positive, the situation is similar.  However, as there is a cosmological horizon,  the mass is also bounded from above as the two horizons merge into each other.  The case of negative cosmological constant is more interesting.  In that case, if the cosmological constant is not too big, there is the possibility of a first-order phase transition between black hole solutions. In fact, if
$$
|\Lambda| >\frac{1}{4r_0^2}\,,
$$
the graph of temperature versus the horizon radius, analog to Fig. \ref{tr} (a) is monotonically increasing, there is no maximal temperature, and all solutions are thermodynamically stable. If, on the contrary,
$$
|\Lambda| <\frac{1}{4r_0^2}\,,
$$
then $T(r_+)$ has two extremal points: $r_+=r_1$ and $r_+=r_2$, as it can be seen in Fig. \ref{tr2} (a).  
The region between those point is thermodynamically unstable, and therefore it has negative heat capacity.

 For small temperatures the system has only one  value for $r_+$. When the temperature is raised to the region $T\in[T_1,T_2]$,  the system has three possible values for $r_+$. For $T\in[T_1,T_0]$, the smallest value of $r_+$ is the one that minimizes the Helmholtz free energy, as it can be seen in Fig.\ref{tr} (b), and therefore that is the value that the system will choose. When $T=T_0$ is reached, the black hole performs a first-order phase transition from point $A$ to point $B$ of Fig.\ref{tr} (a). This is because at that temperature the two stable branches of the Helmholtz free energy intersect each other, and from that point on,  the branch with bigger values of $r_+$ minimizes the free energy. The discontinuity in the slope, as one passes from one branch to the other, is precisely the discontinuity in the entropy characterizing a first-order transition.

\section{Spin from isospin effect and gravitating Merons}
\label{sec:sfi}

In this section we will describe the spin from isospin effect induced on the system of a scalar particle moving in the Meron black hole background. 

Before considering the case of a gravitating Meron, we will briefly review the standard effect in flat space for 't Hooft-Polyakov monopoles found in  \cite{Jackiw:1976xx, Hasenfratz:1976gr, Goldhaber:1976dp}, where a fermionic excitation arises as a ``bound'' system of two bosons: a scalar particle
and the monopole. 

After this, we will move on to the meronic analog of the spin from isospin effect, where
the Meron plays the role of the monopole. It is very important to emphasize that, in this case, the effect is only possible if the
Meron is coupled to gravity (as Merons in flat space are known to be
singular \cite{deAlfaro:1976qet,Callan:1977qs}),  which is the reason why we refer to it as a \emph{gravitational spin from isospin effect}.
Moreover, the singularity of the Meron at $r=0$ is covered by the horizon and thus shielded from the exterior by it. This is a very welcome feature since, otherwise, one would need to regularize such singularity, as it is the case in flat space \cite{Callan:1977qs}. Also, when Yang-Mills theory is coupled to General Relativity, the total mass of the
system has to be computed with the ADM formalism (as described in the previous
subsection), and the result turns out to be finite. These facts are basically the reason why the spin from isospin effect that we will describe in subsection \ref{sec:gravsfi} is a genuine feature of the gravitational interaction.

Finally, in the last part of this section we entertain the possibility of a
holographic realization of this effect, namely, that it should be possible to
compute fermionic observables in the boundary CFT from a purely bosonic theory
in the bulk.

\subsection{Spin from isospin in flat space}
\label{sec:sfiflat}

The idea that the composite of two bosons can become a fermion dates back a long
time \cite{saha1936origin,Saha:1949np} (see also \cite{coleman1983magnetic}). Basically one notes that the total angular momentum of an electric charge in the presence of a magnetic monopole, even in the non-relativistic limit, must include the contribution from the angular momentum stored in the electromagnetic field. Moreover, because of Dirac's quantization condition, this effect contributes with half-integer multiples of $\hbar$, so the quantum mechanical system describing the charge/monopole bound state becomes a fermion.

Consider now the 't Hooft-Polyakov monopole
\begin{equation}
\begin{aligned} \label{eq:tHP-monopole} \Phi^a &= \frac{x^a f(vr)}{r^2}\,,\\ A^a_i &= \frac{\epsilon^a_{ij}x^jg(vr)}{r^2}\,,\\ A^a_0 &= 0 \,,\end{aligned}
\end{equation}
where $f(x)=x\coth(x)-1$, and $g(x)=1-x\,\mathrm{csch}(x)$. Note that this is a four-dimensional example, spatial indices $i,j$ in this case run from $1$ to $3$. The key observation is that this solution does not transform as a vector under spatial
rotations nor under global SU(2) transformations, but it does transform correctly under the
combination
\begin{equation}
J_{a}=\epsilon_{ai}^{\;\;\;j}\,x^{i}\,p_{j}+t_{a}\,, \label{eq:HtH}%
\end{equation}
where $p_{i}$ is the canonical conjugate to $x^{i}$.

Now consider a non-relativistic spin 0 particle represented by the state $|\psi_a\rangle$, where $a$ is an internal isospin index transforming under some irreducible representation of SU(2), interacting with the monopole
background \eqref{eq:tHP-monopole}. Its evolution will be described by a
Schr\"{o}dinger equation $i\partial_{t}|\psi\rangle=H|\psi\rangle$ where $H$
is the Hamiltonian for such particle in the background \eqref{eq:tHP-monopole}. Since this
background transforms as a vector under the $\mathfrak{su(2)}$ generator \eqref{eq:HtH}, we get
\[
\lbrack J_{a},H]=0\,.
\]
Thus, we can diagonalize the Hamiltonian using the eigenstates generated by $J^{a}$. Therefore, since the state representing the monopole, say, $|M\rangle$
and the spinless particle $|\psi_a\rangle$ are both, independently, multiplets
of $SU(2)$, one can think of this scalar-monopole \textquotedblleft bound eigenstate\textquotedblright  $|M\rangle\otimes|\psi_a\rangle$ of $H$ as a state with
total angular momentum $J^{a}$ \cite{Hasenfratz:1976gr}. If the spinless particle $|\psi_a\rangle$ is
in a half-integer representation of the internal SU(2) group, then the bound
state acquires  half-integer spin quantum numbers and, by the spin-statistics connection, it effectively becomes a fermion
\cite{Goldhaber:1976dp}. Moreover, Hasenfratz and
t' Hooft \cite{Hasenfratz:1976gr} also showed that in the classical limit, the angular
momentum stored in the Yang-Mills field above is precisely given by the $t_a$ contribution in \eqref{eq:HtH}.

\subsection{Gravitational spin from isospin}
\label{sec:gravsfi}
Consider a complex scalar (multiplet) field $\Phi$ in the black hole \eqref{eq:Metrica1} and Meron Yang-Mills background \eqref{eq:Ameron}. Its action reads
\begin{equation}
S_{scalar}=-\frac{1}{2}\int d^{5}x\sqrt{-g}\left[  \left(  D_{\mu}\Phi\right)
^{\dag} D^{\mu}\Phi  + m^2|\Phi|^2\right]
\ ,\label{scalaraction}%
\end{equation}
where $D_{\mu}\equiv\partial_{\mu}+A_{\mu}$, and $\Phi$\ is in the fundamental representation of $SU(2)$. Following the logic of \cite{Hasenfratz:1976gr}, let us now imagine quantizing this scalar field around the classical background of the black hole and Meron field. In other words, defining
\eal{
A_\mu &= A_\mu^{{\rm cl}} + A_{\mu}^{{\rm qu}}\,,\\
\Phi &= \Phi^{{\rm qu}}\,,
}
and by only considering the quantum fluctuations $\Phi^{\rm qu}$ (\emph{i.e.,} neglecting $A_{\mu}^{{\rm qu}}$), one would be quantizing the system in the one-soliton sector of the Hilbert space. That is, our excitations correspond to those of the bound-state formed by the scalar and the Meron background.

In the non-relativistic limit, the interaction of this spinless particle (with internal isospin indices) with such background is described by a Hamiltonian of the form
\begin{equation}
H=-\frac{1}{2m}\gamma^{ij}\left(\nabla_i+A_i\right)\left(\nabla_j+A_j\right)\;,\quad i,j=1,2,3,4\,.
\end{equation}
where $\gamma^{ij}$ is the inverse of the spatial metric \eqref{spatial}, and $A_i$ is given by the spatial components of \eqref{eq:Ameron}. Note that the $A_i$'s, and hence $H$, are matrix-valued operators. For simplicity, let us consider the asymptotically flat case $\Lambda =0$. The Hamiltonian has the asymptotic form
\begin{equation}
\label{eq:H-nonrel}
H_{{\rm asymp}}=-\frac{1}{2m}\partial^2_r-\frac{1}{2mr^2}\,h^{mn}D_mD_n\,,
\end{equation}
where $h^{mn}$ denotes the inverse of the metric on the $S^3$ defined in Eq. \eqref{metrics3-2}. In order to simplify some calculations, it will prove useful to express the unit three-sphere in terms of cartesian coordinates $X^M\,,M=1,2,3,4$ such that $\delta_{MN}X^M X^N=1$. In these coordinates the metric on $S^3$ is $h_{MN}=\delta_{MN}$, and the usual angular momentum operator in four spatial dimensions can be simply expressed as
\begin{equation}
\label{ang-mom}
\mathcal{L}_{MN}=-i(x_M \partial_N - x_N \partial_M)\,.
\end{equation}
It is easy to see that this operator does not commute with the Hamiltonian \eqref{eq:H-nonrel}.  Since  partial derivatives transform trivially as vectors under rotations, $r^2$ is rotationally invariant, and in this basis $h^{MN}$ is constant (see Appendix \ref{sec:car}), looking at \eqref{eq:H-nonrel} we see that problem stems from the fact that the gauge connection $A$ does not transform as a vector under rotations generated by \eqref{ang-mom}. Hence, since $\mathcal{L}_{MN}$ is not a conserved quantity, its eigenvalues are not useful to label quantum states. However, the operator
\begin{equation}\label{jtot}
\mathcal{J}_{MN}\equiv \mathcal{L}_{MN}+\mathcal{T}_{MN}\,,
\end{equation}
\emph{does} commute with this Hamiltonian provided $\mathcal{T}_{ab} \equiv \epsilon_{\;ab}^{c}t_c$ and $\mathcal{T}_{4a} \equiv t_a$ (recall that the indices $a,b,c$ run from 1 to 3). Thus, it is $\mathcal{J}_{ij}$ that should be identified with the true angular momentum of the system, and not $\mathcal{L}_{ij}$. This can be seen as follows: expressing the right invariant Maurer-Cartan form \eqref{lmc} in the coordinates $X^M$, the Meron field \eqref{eq:Ameron} takes the form (see Appendix \ref{sec:car})
\begin{equation}
A = A_M dX^M = A_a dX^a + A_4 dX^4\,,
\end{equation}
where\footnote{Even though indices $a,b=1,2,3$ have so far been used as group indices, in \eqref{Acar} they denote spatial indices. This identification between the group manifold and spatial three-sphere is a common ingredient when studying the spin from isospin effect \cite{Jackiw:1976xx}.}
\eal{
A_a = i\left(X^4 t_a - \epsilon_{\;ab}^{c}X^b t_c\right)\,, \qquad
A_4 = -iX_a t_a \label{Acar}\,.
}
Using this one obtains
\begin{eqnarray}
\label{eq:vector}
[\mathcal{J}_{MN},A_R] = i \left(\delta_{MR}A_N-\delta_{NR}A_M\right)\,.
\end{eqnarray}
Hence, $A_M$ rotates as a vector under the action of $\mathcal{J}_{MN}$, and therefore, so does $D_M$. Since the  asymptotic hamiltonian \eqref{eq:H-nonrel} is quadratic in $D_M$, this yields
\begin{equation}\label{chj}
[H_{\rm asymp},\mathcal{J}_{MN}]=0\,,
\end{equation}
as promised.

This analysis can be easily generalized to the asymptotically (A)dS case. In fact, in the presence of cosmological constant, the first term in \eqref{eq:H-nonrel} must be replaced by $\frac{3}{\Lambda m r^2}\partial^2_r\,$, which is rotationally invariant as it only depends on $r^2$. The second term in \eqref{eq:H-nonrel} remains unmodified and the operators \eqref{ang-mom} continue to generate the Lorentz subgroup of the full asymptotic symmetry group. As the Meron has support only on $S^3$, Eq. \eqref{Acar} also remains unchanged, showing that \eqref{chj} holds when $\Lambda\neq 0$.

The operator $\mathcal{L}_{MN}$ can be divided into two sets of angular momenta that satisfy the $\mathfrak{su}(2)$ algebra (for details see Appendix \ref{sec:so4})
\begin{equation}\label{lpandlm}
L^\pm_a=\epsilon_a^{\;bc}\mathcal{L}^{bc}\pm\mathcal{L}_{4a}\,,\quad \left[ L^\pm_a , L^\pm_b \right]=i\epsilon^{c}_{\;ab}L^\pm_c\,.
\end{equation}
In the same way, the total angular momentum $\mathcal{J}_{MN}$ can be divided into two sets of $\mathfrak{su}(2)$ generators, that have the form
\begin{equation}\label{JpJm}
J^{+}_{a}=L^{+}_{a}+t_{a}\, ,\quad J^{-}_{a}=L^{-}_{a}\,.
\end{equation}
The form of $J^{+}$ resembles the shift in \eqref{eq:HtH} and produces the spin from isospin effect.
Thus, in the same way detailed in subsection \ref{sec:sfiflat}, the bound system of a spinless particle in the background of this meronic black hole can acquire
half-integer spin quantum numbers, hence, effectively, becoming a fermion.

It is also instructive to study the Klein-Gordon equation for a scalar field in this background:
\begin{equation}
\left(  \square+\nabla_{\mu}A^{\mu}+2A^{\mu}\nabla_{\mu
}+A^{\mu}A_{\mu}-m^{2}\right)  \Phi=0\label{scalarequ}\,.
\end{equation}
After using \eqref{eq:Metrica1} and \eqref{eq:Ameron}, Eq. (\ref{scalarequ}) becomes

\begin{equation}
\left[ \square-\frac{1}{r^{2}}\left(  4\,\vec{t}\cdot\vec{L}%
^{+}+ \vec{t}\cdot\vec{t} \right) -m^{2}\right]\Phi=0\ ,\label{scalarequ2}%
\end{equation}
where $\vec{t}=(t_{1},t_{2},t_{3})$ and $\vec L^{+}=(L^{+}_{1},L^{+}_{2},L^{+}_{3})$ is the right angular momentum operator defined in \eqref{lpandlm} (for details see Appendix \ref{sec:su2}),
\begin{align}
L_{1}^{+} &  =i\left(\cos\psi\cot\theta\partial_{\psi}+\sin\psi\partial_{\theta
}-\frac{\cos\psi}{\sin\theta}\partial_{\varphi}\right)\, ,\nonumber\\
L_{2}^{+} &  =i\left(\sin\psi\cot\theta\partial_{\psi}-\cos\psi\partial_{\theta
}-\frac{\sin\psi}{\sin\theta}\partial_{\varphi}\right)\, ,\label{rotalgebra}\\
L_{3}^{+} &  =-i\partial_{\psi}\, .\nonumber%
\end{align}
The d'Alembertian on the black hole background \eqref{bhmetric} has the form
\begin{equation}
\square=-\frac{1}{f^{2}}\partial_{t}^{2}+\frac{1}{r^{3}}\partial_{r}\left(r^{3}f^{2}\partial_{r}\right)-\frac{1}{r^2}\mathcal{L}^2\,.
\end{equation}
Here $\mathcal{L}^2$ denotes the square of the four-dimensional angular momentum \eqref{ang-mom} (see Eq. \eqref{L2} in Appendix \ref{sec:so4}). Using this and Eq. \eqref{JpJm}, the equation of motion for the scalar field takes the form 
\begin{equation}\label{scalarequ3}
\left[-\frac{1}{f^{2}}\partial_{t}^{2}+\frac{1}{r^{3}}\partial_{r}\left(r^{3}f^{2}\partial_{r}\right)-\frac{1}{r^2}\left(  \mathcal{J}^{2} - \vec t \cdot \vec t \right)  -m^{2}\right] \Phi=0\,,
\end{equation}
where $\mathcal{J}^2\equiv \frac{1}{2}\mathcal{J}_{MN}\mathcal{J}^{MN}=2\left(J^{+}\right)^2+2\left(J^{-}\right)^2$ is the square of the total angular momentum \eqref{jtot}. Note that the the term $\left(\mathcal{J}^{2} - \vec t \cdot \vec t \right)$ in Eq. \eqref{scalarequ3} is similar to the one found in \cite{Boulware:1976tv}, where the scattering of a scalar particle off a magnetic monopole was studied.

Now we can consider modes of the form
\begin{equation}\label{phimodes}
\Phi_{\ell \m\n}=e^{-i\omega t}D^{\ell}_{\m\n}(\psi,\varphi,\theta)r^{-1}\phi(r) \, ,
\end{equation}
where $D^{\ell}_{\m\n}(\psi,\varphi,\theta)$ stands for the Wigner D-functions \cite{varshalovich1988quantum} (see Eq. \eqref{wdf} in Appendix \ref{sec:su2}), which are the higher-dimensional generalization of the standard spherical harmonics \cite{meremianin2009hyperspherical}. The shift in $L^+\rightarrow J^+$ produces half-integer shift in the angular momentum eigenvalues of the modes $\Phi_{\ell mn}$ given by (see Eq. \eqref{shiftls} in Appendix \ref{sec:so4}),

\begin{equation}\label{shitjs}
\begin{array}{lcl}
\mathcal{J}_{34}  D^\ell_{\m\n}\left(\psi,\theta,\varphi\right)&=& \left(\m-\n+\dfrac{1}{2} \right)  D^\ell_{\m\n}\left(\psi,\theta,\varphi\right) \,,\\[6pt]
\mathcal{J}_{12}  D^\ell_{\m\n}\left(\psi,\theta,\varphi\right)&=& \left(\m+\n+\dfrac{1}{2} \right)  D^\ell_{\m\n}\left(\psi,\theta,\varphi\right)\,.
\end{array}
\end{equation}
As the eigenvalues of the total angular momentum $\mathcal{J}_{MN}$ are given by half-integer multiples of $\hbar$, the scalar field $\Phi$ effectively becomes  fermionic.

\subsection{Holography}
\label{sec:holo}

The gauge/gravity duality states that any theory of quantum gravity that
asymptotically approaches AdS${}_{d+1}$ space-time is dual to a flat
$d$-dimensional conformal field theory on the boundary of AdS. The AdS/CFT
dictionary can be used to relate, in principle, all physical observables of
the gravitational theory to those of the CFT.\footnote{Although it is not yet clear how the boundary theory precisely reproduces the degrees of freedom of the black hole interior (see \cite{Harlow:2018fse} for a recent review on the subject), this will not bother us here since we are dealing with fields in the exterior of the black hole only.} Because of the strong/weak nature of the correspondence, the description of strongly correlated observables in terms of weakly interacting fields in the gravitational bulk is one of the most useful (and widely studied) aspects of the duality.

In the presence of black holes, the bulk partition function is dual to the thermal partition function of the theory on the boundary, where the Hawking temperature of the black hole corresponds to the temperature of a thermal state in
the dual theory. For horizons with spherical topology (such as
the one studied here), the dual $\mathcal{N}=4$ super Yang-Mills
theory is naturally defined on $S^{d-1}\times\mathbb{R}$
space-time.\footnote{However, as noted in \cite{Emparan:1999pm}, depending on
how the radial slices are chosen as one approaches the boundary of
(Euclidean) AdS$_{d+1}$, one could also have the boundary theory defined on other spacetimes such as $S^{d}$, $\mathbb{R}^{d}$, $\mathbb{H}%
^{d}$, and $\mathbb{H}^{d-1}\times\mathbb{R}$.} Since we now have a new black hole solution at our disposal, a natural question is: what can it do for the dynamics of strongly correlated systems?

In this section we argue that the ``gravitational'' spin from isospin effect described in the previous section has the interesting property of describing fermionic observables in the boundary of the spacetime\footnote{We note that this boundary effect is not restricted to the asymptotically AdS case only, but also for asymptotically dS and Minskowski. This could have interesting applications in cosmology (dS) and also in the recent developments of amplitudes on the \emph{celestial sphere}.}, even though the bulk theory is purely bosonic.\footnote{The converse case, \emph{i.e.}, finding the gravitational dual of the usual flat space spin from isospin effect, is also an interesting question which we leave for future work.} We will provide supporting evidence that the composite system of a scalar field and the Meron black hole is dual to a fermionic operator in the boundary CFT. Moreover, we also conjecture that this corresponds to a \emph{double trace} CFT operator.

The first argument is that scalar excitations, in the presence of the Meron black hole, manifest themselves as fermionic degrees of freedom at the AdS boundary. This stems from the fact that the total angular momentum \eqref{jtot} has half-integer quantum numbers. However, there is a more direct way to see this. For our black hole solution \eqref{eq:f-gen} with $\Lambda\equiv -6/L^2<0$, the metric is asymptotically $AdS_5$ and, as $r\to \infty$, we get
\begin{equation}\label{confboundary}
ds^2 \sim\frac{r^2}{L^2}\left(-dt^2 + L^2 h_{mn}dx^m dx^n \right)\,,
\end{equation}
which defines the metric of the conformal boundary, with topology $\mathbb{R}\times S^3$. Consider now the expansion of the scalar field $\Phi$ near the spatial boundary ($r\to \infty$). We look for an Ansatz of the form (see, \emph{e.g.}, \cite{deHaro:2000vlm,Skenderis:2002wp})
\eal{
\label{eq:phi-expansion}
\Phi(r,x)=r^{-\Delta}\left(\varphi^{(0)}(x)+\frac{1}{r^2}\varphi^{(2)}(x)+\cdots \right)
}
where $x$ are the coordinates of the fields that ``live'' on the boundary of $AdS_5$. Since this boundary has the topology $\mathbb{R}\times S^3$, we use $x=(t,\Omega)$ where $\Omega$ collectively refers to the three angular coordinates on the 3-sphere (see equations \eqref{eq:angles} and \eqref{metrics3-2}). Inserting the asymptotic expansion \eqref{eq:phi-expansion} into the scalar wave equation \eqref{scalarequ3} one obtains (at leading order in the large $r$ expansion)
\begin{equation}
\Delta(\Delta-4)-m^2L^2=0\,.
\end{equation}
This is the standard relation between the mass of the scalar field in five dimensions and $\Delta$, which is interpreted as the conformal dimension of the scalar CFT operator dual to $\Phi$. The subleading term yields
\begin{equation}
\label{eq:source-of-source}
\left[-\partial_t^2 -\frac{1}{L^2}(\mathcal{J}^2-\vec t\cdot \vec t)\right]\varphi^{(0)}(t,\Omega)=\frac{4(1-\Delta)}{L^4} \varphi^{(2)}(t,\Omega)\,.
\end{equation}
In the absence of the Meron, $\mathcal J^2=\mathcal{L}^2=-\nabla^2_{S^3}$ and $\vec t=0$, thus $\varphi^{(0)}$ would represent a massless scalar field in four spacetime dimensions moving on a 3-sphere coupled to the source $\varphi^{(2)}(t,\Omega)$. Turning on the Meron gauge field in the bulk, however, yields equation \eqref{eq:source-of-source} with the total angular momentum $\mathcal{J}$ acquiring an extra term (see \eqref{jtot}) due to the coupling between the isospin indices of $\Phi$ and the Meron. We can re-write the equation above as
\begin{equation}\label{KGrxs3}
\left[-\partial_{t}^{2}+\frac{1}{L^2}h^{mn}\left(\nabla_{m}+A_{m}\right)\left(\partial_{n}+A_{n}\right)\right]\varphi^{(0)}=\frac{4(1-\Delta)}{L^4} \varphi^{(2)}\,,
\end{equation}
where  $\nabla_{m}$ is the covariant derivative with respect
to the affine connection on $S^{3}$, and $A_m$ denotes the angular components of the Meron gauge field \eqref{eq:Ameron}\footnote{Equation \eqref{KGrxs3} also appears in the description of scalar fields with generalized Wentzell boundary conditions \cite{Zahn:2015due,Dappiaggi:2018pju}.}. Due to the topological nature of our Meron $A_\mu$ (which extends all the way to infinity), and because we are studying the scalar field in the probe limit, one can consider the Meron as an external field which couples to the boundary degrees of freedom of the scalar (isospin doublet) $\Phi$.

Now we will see that equation \eqref{KGrxs3} can actually be reinterpreted as describing fermionic degrees of freedom.  In order to do this, let us consider a Majorana fermion in the chiral representation,
\begin{equation}\label{majoranafermion}
\Psi=\left(\begin{array}{c}
\psi\\
-i\sigma_{2}\psi^{*}
\end{array}\right)\,,
\end{equation}
defined on $\mathbb{R}\times S^3$ (see for instance,  \cite{Pal:2010ih,Dvornikov:2011bu,henn2014scattering}). This, in turn, can be described in terms of the two-spinor $\psi$ satisfying a massive Weyl equation (see Appendix \ref{sec:weyl}). Adding a source term $\rho_\psi$ for $\psi$, the equation becomes
\begin{equation}
\partial_{t}\psi-e^m_a\sigma^{a}\left(\partial_{m}+\Gamma_{m}\right)\psi+\mu\,\sigma_{2}\psi^{*}=-\rho_\psi\label{eq:weyleq}\,,
\end{equation}
where $\sigma^a$ are the Pauli matrices, $e_{a}^{m}$ denote the inverse dreibein associated to the metric \eqref{metrics3-2}, and $\Gamma_m$ is the spin connection on $S^3$. If we consider the corresponding Klein-Gordon equation for the two-spinor $\psi$ (see Eq. \eqref{KGweyl}), one obtains
\begin{equation}\label{KGrxs32}
\left[-\partial_{t}^{2}+\frac{1}{L^2}h^{mn}\left(\nabla_{m}+\Gamma_{m}\right)\left(\partial_{n}+\Gamma_{n}\right)+\frac{3}{2L^2}-\mu^{2}\right]\psi=-\rho \,.
\end{equation}
where the right hand side of Eq. \eqref{KGrxs32} can be written in terms of $\rho_\psi$ and its derivatives, but the explicit form is not important here. Notice that equation \eqref{KGrxs32} does not make any reference to a gauge field coupling.  Nevertheless, using Eqs. \eqref{dreibeins3} and \eqref{spincon}, it is straightforward to see that, on $S^3$, the components of the spin connection $\Gamma _m$ match the angular components of the Meron gauge field \eqref{merona1} (for $\lambda=1/2$)
\begin{equation}\label{match}
\Gamma_m \, dx^m=\frac{i}{4}\omega_R^a \sigma _a = A_m dx^m\,.
\end{equation}
Thus, \eqref{KGrxs32} is on-shell equivalent to \eqref{KGrxs3} when $\psi$ is identified with $\varphi^{(0)}$, the source term $\rho$ is identified with $\frac{4(1-\Delta)}{L^4}\varphi^{(2)}$, and  the mass of the Majorana fermion is fixed to $\mu^2=3/2L^2$. This means that the original scalar excitations of an isospin doublet $\Phi$, coupled to the Meron gauge field through its isospin indices, can be reinterpreted as a two-spinor $\psi$ with fixed mass, coupled to the geometry of the $AdS_5$ conformal boundary and satisfying \eqref{KGrxs32}. Hence, knowing a solution of \eqref{KGrxs3}, allows us to construct a Majorana spinor at the boundary by means of \eqref{majoranafermion} by reinterpreting isospin indices of $\Phi$ as spin indices. This argument is another way to justify why the Wigner D-functions in the modes \eqref{phimodes} must be chosen as fermionic, pretty much in the same way as it happens when describing fermions in terms of the Skyme model \cite{Adkins:1983ya} (see also \cite{Rho:1994ni}). All this put together signals that there must be a way to construct a holographic dictionary between a purely bosonic theory in the bulk and a fermionc theory at the boundary using the spin from isospin effect. It is important to note that, even though we expect the Meron-scalar bound state to be a fermion, the fact that the probe scalar field has no back reaction means we are considering its mass to be very small compared to the mass of the Meron black hole. Thus, in this limit, the properties of the fermionic bound state can be approximated by the dynamics of the scalar field only, which is tantamount to studying the properties of the hydrogen atom in the limit when the proton is fixed at the origin because it is much heavier than the electron.

A stronger argument in support of this idea is the following. The GKPW dictionary \cite{Gubser:1998bc,Witten:1998qj} states that the asymptotic value of $\Phi(x,r)$ at the AdS boundary ($r \to \infty$) acts as the source for a scalar operator in the CFT. Since $\Phi$ is a bulk scalar, its asymptotic value is dual to a scalar operator $\mathcal{O}^a$ in the 4-dimensional CFT \cite{Gubser:1998bc,Witten:1998qj}. Near the boundary ($r\to \infty$) the scalar field behaves as given in \eqref{eq:phi-expansion}. It is standard to regard $\varphi^{(0)}$ as the source for the operators $\mathcal{O}^a$ of the CFT, which in turn yields $\varphi^{(2)}$ to be directly proportional to their vacuum expectation value $\langle \mathcal{O}^a\rangle$. From \eqref{scalarequ3} it is easy to see that, at leading order, the asymptotic behavior of the scalar field near the boundary is insensitive to the presence of the gauge field $A_\mu$. This is simply because the term that accompanies $\mathcal{L}^2$ in the Klein-Gordon equation, which gets modified as 
$$
\mathcal{L}^2 \to \mathcal{J}^2 - \vec t \cdot \vec t\,,
$$ 
is subleading in the large $r$ expansion. Thus, one may wonder in what sense the ``spin from isospin'' effect modifies the CFT observables. It has, however, an immediate effect. In order to see this, it is useful to use the identification between two Fock spaces: the space of (spinless) free particle states in AdS and the space of states in the CFT \cite{Witten:1998qj,Balasubramanian:1998de,Balasubramanian:1998sn,Balasubramanian:1999ri} (see also \cite{Fitzpatrick:2010zm,Fitzpatrick:2011jn}).\footnote{At leading order in the $1/N$ expansion in the CFT, this identification is exact. This is indeed all we need since subleading corrections in $1/N$ correspond to stringy corrections to the effective gravitational theory which we are not considering here.} In the $AdS_5$ description, a free scalar field $\Phi$ can be written as 
\eal{
\Phi= \sum_{n,\ell,\m,\n} \Phi^{n}_{\ell \m\n} \, a^{n}_{\ell \m\n} + \Phi^{n\,*}_{\ell \m\n} \, a^{n\,\dagger}_{\ell \m\n}\,,
}    
where the modes $\Phi^{n}_{\ell \m\n}$ satisfy the wave equation \eqref{scalarequ3}, $n$ is their energy eigenvalue, $\ell$ is the total angular momentum quantum number, while $\m,\n$ are the ``magnetic'' angular momentum quantum numbers that arise from the $S^3$ of $AdS_5$ (\emph{e.g.}, in $AdS_4$, they would simply be the usual ``$m$'' eigenvalue of $L_z$ of three spatial dimensions). Because of our hedgehog-type solution for the gauge
field \eqref{eq:Ameron}, the CFT at the boundary of $AdS_5$ is
naturally defined on $S^{3}$. Now, the boundary operator $\mathcal{O}$, dual to $\Phi$, inherits a similar decomposition which, in Lorentzian radial quantization, has the form
\eal{
\label{eq:op-expansion}
\mathcal{O}(t,\Omega) = \sum_{n,\ell,\m,\n} \frac{1}{ N_{n \ell \m\n}^{\mathcal{O}}}  \left[e^{i(2\Delta+2n+\ell)t} Y^{\ell}_{\m\n}(\Omega) a^{n}_{\ell\m\n} + e^{-i(2n+\ell)t} Y^{\ell \,*}_{\m\n}(\Omega) a^{n\,\dagger}_{\ell \m \n}\right]\,,
}
where $(t,\Omega)$ are the spacetime coordinates of the 4-dimensional CFT, $\Delta$ is the conformal dimension of the operator $\mathcal{O}$, $N_{n \ell \m\n}^{\mathcal{O}}$ is a normalization factor, and $Y^{\ell}_{\m\n}(\Omega)$ are the spherical harmonics on $S^3$. The crucial point is the following: as discussed in Section 4.2, turning on the Meron field in the bulk shifts the eigenvalues of the total angular momentum $\mathcal J$ by \emph{half-integer numbers} as in \eqref{shitjs} (provided the scalar field is in one of the half-integer representations of SU(2) of isospin). Therefore, the spherical harmonics $Y^{\ell}_{\m\n}$ in \eqref{eq:op-expansion}, which (for $AdS_5$) are just the Wigner D-functions \eqref{phimodes}, will also see its angular momentum quantum numbers ($m_1,m_2$) shifted by $\pm 1/2$ (see Eq. \eqref{shitjs}), making the putative scalar $\mathcal{O}$ in \eqref{eq:op-expansion} now a fermionic operator! \footnote{As a consequence, there will be no \textquotedblleft s-wave\textquotedblright\ since the eigenvalues of $\mathcal{J}$ would never vanish.}

At leading order in the $1/N$ expansion, the CFT has an exact Fock space representation given by the creation and annihilation operators $a^{n\,\dagger}_{\ell\m\n}$ and $a^{n}_{\ell\m\n}$ in \eqref{eq:op-expansion}. From the point of view of AdS physics, these operators play a central role: they correspond to the creation and annihilation of particle states in $AdS_5$ with definite energy $\omega_{n,\ell}=\Delta+2n+\ell$ and angular momentum quantum numbers $(\ell,\m,\n)$. Their CFT duals are primary states (and their descendants). For instance, a one-particle state in AdS with mininum energy $\omega_{0,0}=\Delta$ corresponds to a {\it primary} state in the CFT, while excited states in AdS correspond to descendent states. Multi-particle states in AdS are dual to CFT multi-trace operators, in particular, a two-particle state in AdS is dual to a double-trace operator. Their conformal dimension is just the sum of the dimensions of the CFT primaries simply because the energy of a system of free particles in AdS is the sum of the individual energies of its constituents. However, when interactions among the particles are turned on, the total energy of the multi-particle state will change. In particular, when a bound state of two particles is formed, its total energy will be different to that of the sum of the individual particles. In our case, we have a bound state of a scalar particle and the gravitating Meron which, altogether, behaves as a fermion. Thus, we expect that this new system, originally formed out of two bosons, describes a fermionic double trace operator in the CFT \cite{Witten:2001ua}, whose conformal dimension differs from that of the sum of CFT primaries.

If we would like to go beyond this and start considering a self-interacting scalar,\footnote{To compute, for instance, higher than 2-point correlation functions} all correlation functions can be computed, in principle, from the action \eqref{scalaraction} by adding a corresponding potential $U(\Phi)$, and using the standard AdS/CFT dictionary.\footnote{See, for instance, Section 5 in \cite{Skenderis:2002wp}.} Moreover, to obtain
real-time Green's functions one needs to solve the equation of motion for free
fields in the gravitational theory with suitable boundary conditions at
both, the AdS boundary, and deep in its interior. When black holes are
present, \emph{retarded} Green's functions in the CFT are obtained by keeping the infalling modes only at the black hole horizon
\cite{Son:2002sd,Herzog:2002pc,Marolf:2004fy,Gubser:2008sz,Skenderis:2008dg,Iqbal:2008by}, while setting the outgoing ones to zero. At the boundary of AdS the necessary boundary conditions are those that deem
the fields normalizable \cite{Gubser:1998bc,Witten:1998qj}. For fermionic fields, however, there are some subtleties which are not present
for bosonic ones. More specifically, the boundary behavior of the Dirac fields,
together with the necessity of having a well-defined variational principle
\cite{Henneaux:1998ch} needs a more subtle treatment compared to that of scalars\footnote{The reason simply stems from the fact that free fermion
fields satisfy first-order differential equations. See section III in \cite{Iqbal:2009fd} for details.} (see also \cite{Henningson:1998cd,Mueck:1998iz}). Thus,
we believe that being able to compute fermionic CFT correlators in this alternative way, using only bosonic fields in the bulk, would be a useful addition to the
AdS/CFT dictionary. Recall, however, that it is the scalar-Meron composite that can acquire half-integer spin quantum numbers, not the individual fields. Thus, we need to construct this composite field and solve its wave equation near the AdS boundary in order to obtain the corresponding source for the dual CFT fermionic operator. It is not immediately clear how to derive such a wave equation since, even for the simplest \emph{spin from isospin effect} (the one with a scalar and a t' Hooft-Polyakov monopole), there are many subtleties involved \cite{Goldhaber:1976dp}. We believe, however, that this is a very interesting direction which we leave as future work. It is important to note that in section 4.3 we were able to do so because the effect was already visible at the level of the equation of motion for the scalar field because we were dealing with a free (non self interacting) scalar field.



\section{Conclusions and outlook}
\label{sec:conclusions}

We have constructed an analytic black hole solution of the Einstein-Yang-Mills theory in
five dimensions in which the gauge field is of Meron-type: it is
proportional to a pure gauge, and it has a
non-trivial topological charge. Our solution, referred to as "meronic black hole", has only one integration constant, namely, its mass. The thermodynamics of the meronic black hole is similar to that of Reissner-N{\"o}rdstrom. Here, however, the analog of the electric charge is the Yang-Mills coupling,  which is not a thermodynamical variable, but a fundamental constant, and therefore its variation does not appear in the first law.  In the case of a negative cosmological constant, there is a first-order phase transition when the cosmological scale is big compared to the one defined by the Yang-Mills coupling in Eq. \eqref{l2}.

We would like to point out that black holes with a similar spacetime metric, but with an $SO(4)$ Wu-Yang-like field, have been reported in \cite{Brihaye:2002jg,Mazharimousavi:2008ap,Bostani:2009zf}. Also, a black hole solution with a logarithmic term in the blackening factor (referred to as $f^2$ here) and an $SU(2)$ gauge field has been studied in \cite{Okuyama:2002mh,Brihaye:2007jua}. Although such solution is similar to the one reported here, one may show that it is different from ours, for its topological charge \eqref{chern} vanishes. 

A non-trivial effect of the present solution is that the total angular momentum of the composite system of a spinless particle (with internal isospin quantum numbers) and the Meron black hole, presents half-integer eigenvalues, thus, describing a fermion. Such a shift is a peculiar fingerprint of the non-trivial topological
character of the meronic black hole. In fact, as the Meron singularity is hidden by the black hole horizon, this effect is only possible due to gravity. Because of this reason we have referred to it as a \textit{gravitational spin from isospin} effect. 

In the context of AdS/CFT it is convenient to consider the Klein Gordon equation for a scalar isospin doublet in the Meron black hole background, and study its behavior at spatial infinity. The leading term gives the standard relation between the mass of the scalar and the conformal dimension of the dual CFT operator. At subleading order, a Klein-Gordon equation on the AdS boundary is obtained. This equation is the same that one obtains by ``squaring'' the free Weyl equation for a massive two-spinor on the three-sphere. No gauge field appears here whatsoever. Therefore, the minimal coupling of the scalar isospin doublet to the Meron gauge field, plays the role of the connection of a free spinor to the curvature of the 3-sphere near the $AdS_5$ boundary. As a consequence it seems that, even the \emph{title} of a boson or a fermion to a given boundary operator must take into proper
account the topological nature of the bulk fields. Also, because the radial coordinate in AdS corresponds to the scale of renormalization group in the CFT, the logarithmic dependence in $r$ that we obtained (see Eq. \eqref{eq:f-gen}) seems to indicate that dual theory must also have a logarithmic RG flow like the ones found in \cite{Klebanov:1999rd,Klebanov:2000hb}.

Finally, since bound states in the bulk are dual to multi-trace operators in the boundary CFT, we argued that the bound state of the Meron+Scalar system describes a fermionic double trace operator in the CFT. An important test would be to check if the boundary properties of the putative composite field match with those of a double-trace operator in the 4-dimensional CFT. We are currently working on this and we hope to include it in a upcoming article.    
\section*{Acknowledgements}

We thank Ignacio Araya, Claudio Bunster, and Alberto Faraggi and for very helpful discussions. This work has been supported by FONDECYT grants 1160137, 3160581, and 11171148. Seung Hun Oh is supported by the National Research Foundation of Korea funded by the Ministry of
Education of Korea (Grant 2018-R1D1A1B0-7048945). The Centro de Estudios
Cient\'ificos (CECs) is funded by the Chilean Government through the Centers
of Excellence Base Financing Program of CONICYT. P. Salgado-Rebolledo acknowledges DI-VRIEA for financial support through Proyecto Postdoctorado 2018 VRIEA-PUCV.

\appendix

\section{SU(2)}
\label{sec:su2}

In the fundamental representation, the generators of the $\mathfrak{su}(2)$ algebra are given by $t_a=\frac{\sigma_a}{2}$, $a=1,2,3$, where $\sigma _a$ denotes the Pauli matrices
\begin{equation}\label{paulimatrices}
\sigma_1 = \begin{pmatrix}
    0 & 1\\
    1 & 0 
\end{pmatrix}
\, , \quad 
\sigma_2 = \begin{pmatrix}
    0 & -i\\
    i & 0 
\end{pmatrix}
\, , \quad 
\sigma_3 = \begin{pmatrix}
    1 & 0\\
    0 & -1 
\end{pmatrix}\,.
\end{equation}
They satisfy the relations
\begin{equation}\label{paulirel}
\lbrack t_{a},t_{b}]=i\epsilon_{\;ab}^{c}t_{c}\, ,\quad\mathrm{Tr}[t_{a}%
t_{b}]=\frac{1}{2}\delta_{ab} \, , \quad t_a t_b = \frac{1}{4}\delta_{ab} \mathbf{I} + \frac{i}{2}\epsilon_{abc}t_{c}\,.
\end{equation}

The elements of the group $SU(2)$ can be parametrized in terms of the Euler variables
\begin{equation}\label{ueuler}
U\left(\psi,\theta,\varphi\right)= e^{-i\psi t_3}e^{-i\theta t_2}e^{-i\varphi t_3}\,,
\end{equation}
where $0\leq\psi<4\pi\,,\,\,0\leq\theta<\pi\,,\,\,0\leq\varphi<2\pi$.

The right and left invariant one-forms on SU(2) are given by
\begin{equation}\label{dualforms}
dUU^{-1}=-i\omega^a _R t_a\,, \quad\quad U^{-1}dU=-i\omega^a _L t_a\,.
\end{equation}
and satisfy the Maurer-Cartan equations
\begin{equation}
d \omega_{R}^{a}=\frac{1}{2}\epsilon^{a}_{\;bc}\omega_{R}^{b}\wedge\omega_{R}^{c} \, , \quad d \omega_{L}^{a}=-\frac{1}{2}\epsilon^{a}_{\;bc}\omega_{L}^{b}\wedge\omega_{L}^{c}\,.
\end{equation}

Using \eqref{ueuler}, the components of the Maurer-Cartan forms \eqref{dualforms} read
\begin{align}
&\omega_{R}^{1}=-\mathrm{sin}\psi d\theta+\mathrm{cos}\psi\,
\mathrm{sin}\theta d\varphi\,, \quad\quad\quad && \omega_{L}^{1} =\mathrm{sin}\varphi d\theta-\mathrm{sin}\theta
\mathrm{cos}\varphi d\psi \, , \nonumber\\
&\omega_{R}^{2}=\mathrm{cos}\psi d\theta+\mathrm{sin}\psi\,
\mathrm{sin}\theta d\varphi \,, \quad\quad\quad  && \omega_{L}^{2} =\mathrm{cos}\varphi d\theta+\mathrm{sin}\theta
\mathrm{sin}\varphi d\psi\,, \label{rlmc}\\
&\omega_{R}^{3}=d\psi+\mathrm{cos}\theta d\varphi\,, \quad\quad\quad  && \omega_{L}^{3}  =\mathrm{cos}\theta d\psi+d\varphi\,. \nonumber
\end{align}%
The group $SU(2)$ is diffeomorphic to $S^{3}$, which means that the
bi-invariant metric on $SU(2)$ corresponds to a metric on the three-sphere \cite{milnor1977fundamental} . Therefore $d\Omega^{2}_{S^{3}}$ can be written in the form
\begin{equation}
\label{metrics3bi}
d\Omega^{2}_{S^{3}}=-\frac{1}{2}\mathrm{Tr}[dUU^{-1}\otimes
dUU^{-1}]=-\frac{1}{2}\mathrm{Tr}[U^{-1}dU\otimes
U^{-1}dU]\,.
\end{equation}
Using \eqref{rlmc}, it takes the form
\begin{equation}
d\Omega_{S^{3}}^{2}=h_{mn}dx^{m}dx^{n}\;,\quad h_{mn}=\frac{1}{4}\left(\begin{array}{ccc}
1 & 0 & \cos\theta\\
0 & 1\\
\cos\theta & 0 & 1
\end{array}\right) \; , \quad x^{m}=\left\{ \psi,\theta,\phi\right\} \,. \label{metrics3-3}%
\end{equation}
The left and right invariant vector fields on $SU(2)$ are given by
\begin{align}
&\xi^{R}_{1} =-\cos\psi\cot\theta\partial_{\psi}-\sin\psi\partial_{\theta
}+\frac
{\cos\psi}{\sin\theta}\partial_{\varphi}\,, \quad && \xi^{L}_{1}  =\cos\varphi\cot\theta\partial_{\varphi}-\frac
{\cos\varphi}{\sin\theta}\partial_{\psi}+\sin\varphi\partial_{\theta
}\,, \nonumber\\
&\xi^{R}_{2} =-\sin\psi\cot\theta\partial_{\psi}+\cos\psi\partial_{\theta
}+\frac
{\sin\psi}{\sin\theta}\partial_{\varphi} \,, \quad&& \xi^{L}_{2}  =-\sin\varphi\cot\theta\partial_{\varphi}+\frac
{\sin\varphi}{\sin\theta}\partial_{\psi}+\cos\varphi\partial_{\theta
} \,, \nonumber \\
&\xi^{R}_{3} =\partial_{\psi} \,, \quad  && \xi^{L}_{3}  =\partial_{\varphi} \,. \label{rlvf}
\end{align}
They satisfy
\begin{equation}
[\xi_{a}^{R},\xi_{b}^{R%
}]=-\epsilon_{\;ab}^{c}\xi_{c}^{R} \, , \quad 
[\xi_{a}^{L},\xi_{b}^{L%
}]=\epsilon_{\;ab}^{c}\xi_{c}^{L} \,,
\end{equation}
and are dual to \eqref{lmc}, \emph{i.e.}, $\xi
_{a}^{R}(\omega_{R}^{b})=\delta_{a}^{b}=\xi
_{a}^{L}(\omega_{L}^{b})$. They can be used to define the Laplacian on the 3-sphere:
\begin{equation}\label{l3s}
\frac{1}{4}\nabla^{2}_{S^{3}}=\vec\xi_R \cdot \vec\xi_R=  \vec\xi_L \cdot \vec\xi_L = \frac{1}{\sin^2 \theta} \left( \frac{\p^2}{\p \psi^2}+ \frac{\p^2}{\p \varphi^2} -2\cos\theta\frac{\p^2}{\p\psi \p\varphi}   \right) +\frac{\p^2}{\p \theta^2}+\cot\theta \frac{\p}{\p \theta}  \,.
\end{equation}
where we have adopted the notation $ \vec\xi\cdot\vec\xi\equiv\delta^{ab}\xi_a\xi_b$.

The definition \eqref{rlvf} allows one to define two sets of angular momentum operators:
\begin{equation}\label{LLb}
L^{+}_a = -i \xi_{a}^{R} \, , \quad L^{-}_a = i \xi_{a}^{L} \, , \quad [L^{\pm}_{a},L^{\pm}_{b}]= i\epsilon_{\;ab}^{c}L^{\pm}_{c}\,.
\end{equation}

The operators $\left( L^{\pm}\right)^2\equiv \vec L^{\pm} \cdot \vec L^{\pm}$, and $L^{\pm}_3$ have common eigenfunctions given by the Wigner D-functions $D^\ell_{\m\n}\left(\psi,\theta,\varphi\right)$ \cite{varshalovich1988quantum}, which are equivalent to the hyperspherical harmonics on $S^3$ \cite{meremianin2009hyperspherical},
\begin{equation}\label{wdf}
D^\ell_{\m\n}\left(\psi,\theta,\varphi\right) = e^{-i\m \psi} d^\ell_{\m\n}\left(\theta\right) e^{-i\n \varphi}\,,
\end{equation}
where 
\begin{equation}\label{wdf2}
d^\ell_{\m\n}\left(\theta\right) = \left| (\ell+\m)!(\ell-\m)!(\ell+\n)!(\ell-\n)! \right|^{1/2} \sum_k \frac{(-1)^k  \left(\cos\frac{\theta}{2}\right)^{2\ell-2k+\m-\n} \left(\sin\frac{\theta}{2}\right)^{2k-\m+\n}}{\k!(\ell+\m-k)!(\ell-\n-k)!(\n-\m+k)!}\,.
\end{equation}
In fact, they satisfy
\begin{align}
\left( L^{\pm}\right)^2 D^\ell_{\m\n}\left(\psi,\theta,\varphi\right) &= \ell\left(\ell+1\right)\, D^\ell_{\m\n}\left(\psi,\theta,\varphi\right) \, ,\nonumber\\
L^{+}_3 D^\ell_{\m\n}\left(\psi,\theta,\varphi\right)  &  =-\m \, D^\ell_{\m\n}\left(\psi,\theta,\varphi\right)\, ,\nonumber\\
L^{-}_3 D^\ell_{\m\n}\left(\psi,\theta,\varphi\right)  &  = \n \, D^\ell_{\m\n}\left(\psi,\theta,\varphi\right)\, .
\end{align}

\section{SO(4)}
\label{sec:so4}

The four dimensional analog of the usual three-dimensional angular momentum is a tensor $\mathcal{L}_{MN}\,,\; M,N=1,2,3,4,\, $ which satisfies:
\begin{equation}\label{so4algebra}
[\mathcal{L}_{MN},\mathcal{L}_{RS}]=i\left(\delta_{MR}\mathcal{L}_{NS}+\delta_{NS}\mathcal{L}_{MR}-\delta_{MS}\mathcal{L}_{NR}-\delta_{NR}\mathcal{L}_{MS}\right)\,,
\end{equation}
This corresponds to the $\mathfrak{so}(4)$ algebra, which is isomorphic to $\mathfrak{su}(2)\oplus\mathfrak{su}(2)$. In fact, by defining new generators
\begin{eqnarray}\label{so4}
M_{c}=\tfrac{1}{2}\epsilon_{\,\,\,\,\,\,c}^{ab}\mathcal{L}_{ab}  \, , & P_{a}=\mathcal{L}_{4a}\,,\quad {a,b,c=1,2,3,}
\end{eqnarray}
the algebra \eqref{so4} can be written as
\begin{equation}
\left[M_{a},M_{b}\right] = i\epsilon_{\;ab}^{c}M_{c}\,,\quad
\left[M_{a},P_{b}\right] = i\epsilon_{\;ab}^{c}P_{c}\,,\quad
\left[P_{a},P_{b}\right] = i\epsilon_{\;ab}^{c}M_{c}\,.
\end{equation}
Then, the redefinition
\begin{equation}\label{ls}
L^{\pm}_{a}=\frac{1}{2}(M_{a}\pm P_{a})\, ,
\end{equation}
leads to two copies of the $\mathfrak{su}(2)$ algebra 
\begin{equation}
[L^{\pm}_{a},L^{\pm}_{b}]= i\epsilon_{\;ab}^{c}L^{\pm}_{c} \, , \quad [L^{+}_{a},L^{-}_{b}]=0\,.
\end{equation}
The square of the angular momentum operator in the different basis reads
\begin{equation}\label{L2}
\mathcal{L}^2=\vec{M}\cdot\vec{M}+\vec{P}\cdot\vec{P}= 2\left(L^{+}\right)^2+2\left(L^{-}\right)^2=4 \left(L_{\pm}\right)^2= -\nabla^{2}_{S^{3}}\, ,
\end{equation}
where $\mathcal{L}^2\equiv \frac{1}{2}\mathcal{L}_{MN}\mathcal{L}^{MN}$ and $\nabla^{2}_{S^{3}}$ is the Laplacian of the three-spehere defined in \eqref{l3s}. We can identify the $\mathfrak{su}(2)$ generators $L^{\pm}_a$ defined in \eqref{ls} with the right and left angular momentum operators defined in \eqref{LLb} in terms of right and left invariant vector fields. This means that the Wigner D-functions \eqref{wdf} are also eigenfunctions of $\mathcal{L}^2$
\begin{equation}
\mathcal{L}^2  D^\ell_{\m\n}\left(\psi,\theta,\varphi\right) =l(l+2) D^\ell_{\m\n}\left(\psi,\theta,\varphi\right)  \,,\quad l=2\ell=0,1,2,\dots,
\end{equation}
which agrees with the fact that $\nabla^{2}_{S^{d}}$ has eigenvalues $l(1-d-l)$ when acting on higher-dimensional spherical harmonics \cite{Frye:2012jj}. At the same time, as the Wigner D-functions are eingenfunctions of $L^{\pm}_3$, we obtain the relations
\begin{equation}\label{shiftls}
\begin{array}{lcl}
\mathcal{L}_{34}  D^\ell_{\m\n}\left(\psi,\theta,\varphi\right)&=& \left(\m-\n \right)  D^\ell_{\m\n}\left(\psi,\theta,\varphi\right) \,,\\[6pt]
\mathcal{L}_{12}  D^\ell_{\m\n}\left(\psi,\theta,\varphi\right)&=& \left(\m+\n \right)  D^\ell_{\m\n}\left(\psi,\theta,\varphi\right)\,.
\end{array}
\end{equation}

\section{Cartesian coordinates}
\label{sec:car}
Using Cartesian coordinates $X^M,\, M=1,2,3,4$, the unit three-sphere can be easily described. The metric on $S^3$ has the simple form
\begin{equation}\label{metrics3-4}
d\Omega_{S^3} ^2=\delta_{MN} \; dX^M dX^N\,,\quad\quad\delta_{MN}X^M  X^N =1\;.
\end{equation}
and its isometry group is generated by the angular momentum operator
\begin{equation}\label{lmn}
\mathcal{L}_{MN}=-i\left(X_M\partial_N - X_N\partial_M \right)\,,
\end{equation}
which satisfy the $\mathfrak{so}(4)$ algebra \eqref{so4algebra}. Following \eqref{ls}, one can use \eqref{lmn} to define two sets of $\mathfrak{su}(2)$ vector fields:
\begin{equation}\label{lmmcar}
L_a^\pm = -\frac{i}{2}\left(\epsilon^{c}_{\; ab}X^{b}\partial_{c} \pm X^{4}\partial_{a} \mp X_{a}\partial_{4}\right)\,, 
\end{equation}
from which one can read off the corresponding dual one-forms \eqref{dualforms},
\begin{equation}\label{rlmccar}
\begin{array}{lcl}
dUU^{-1} &=& -2i\left(\epsilon^{a}_{\;bc}X^{b}dX^{c} + X^{4}dX^{a} - X^{a}dX^{4}\right)t_a \,,\\[6pt]
U^{-1}dU &=& -2i\left(\epsilon^{a}_{\;bc}X^{b}dX^{c} - X^{4}dX^{a} + X^{a}dX^{4}\right)t_a\,.
\end{array}
\end{equation} 
The three-sphere embedded in the complex plane $\mathbb{C}^2$ can be parametrized by means of Hopf coordinates $\left(\zeta_1 , \; \zeta_2 , \; \chi\right)$ :
\begin{equation}
Z_1=e^{i\zeta_1}\sin{\chi}\;,\quad\quad Z_2=e^{i\zeta_2}\cos{\chi}\,.
\end{equation}
They are related to the Euler variables introduced in \eqref{ueuler} by $\zeta_1=(\varphi-\psi)/2,\; \zeta_2=(\varphi+\psi)/2,\;\chi=\theta/2$, and allow one to define Cartesian coordinates for $S^3$ by defining $Z_1=iX_1+X_2,\;Z_2=iX_3+X_4$, which lead to
\begin{equation}\label{xs}
\begin{array}{lcl}
&X_1=\sin{\dfrac{\theta}{2}}\sin{\dfrac{\varphi-\psi}{2}}\,,\quad\quad X_2=\sin{\dfrac{\theta}{2}}\cos{\dfrac{\varphi-\psi}{2}}\,.\\[6pt]
&X_3=\cos{\dfrac{\theta}{2}}\sin{\dfrac{\varphi+\psi}{2}}\,,\quad\quad X_4=\cos{\dfrac{\theta}{2}}\cos{\dfrac{\varphi+\psi}{2}}\,.
\end{array}
\end{equation}
It is straightforward to check that , using \eqref{xs}, the metric \eqref{metrics3-4} is brought to \eqref{metrics3-3}, while the $\mathfrak{su}(2)$ operators \eqref{lmmcar} reduce to \eqref{LLb} in terms of \eqref{rlvf}, and the components of the Maurer-Cartan forms \eqref{rlmccar} reduce to \eqref{rlmc}.

\section{Massive Weyl spinor on $\mathbb{R}\times S^{3}$}
\label{sec:weyl}

A Majorana 4-spinor $\Psi$ in four-dimensional Minkowski space can be written in the usual chiral representation as
\begin{equation}\label{majorana}
\Psi=\left(\begin{array}{c}
\psi\\
-i\sigma_{2}\psi^{*}
\end{array}\right)\,,
\end{equation}
where $\psi$ is a two-spinor satisfying a massive Weyl equation \cite{Pal:2010ih,Dvornikov:2011bu,henn2014scattering}
\begin{equation}\label{eq:weyleq}
\partial_{t}\psi-\sigma^{i}\partial_{i}\psi+\mu\,\sigma_{2}\psi^{*}=0\,, 
\end{equation}
and $\mu$ denotes the mass of $\Psi$. Now we want to generalize this description to the case a Majorana fermion field defined on the conformal boundary of $AdS_{5}$ \eqref{confboundary}. Therefore, the metric has the topology of $\mathbb{R}\times S^{3}$ and it can be written in the form
\begin{equation}\label{metricrxs3}
ds_{\mathbb{R}\times S^{3}}^{2}=-dt^{2}+L^{2}\,h_{mn}dx^{m}dx^{n}\,,
\end{equation}
where $L^2 \,h_{mn}$ denotes the metric on a  three-sphere of raduis $L=-6/\Lambda$ (see Eq. \eqref{metrics3-3}). As we are dealing with curved spatial section, Eq. \eqref{eq:weyleq} must be generalized to
\begin{equation}
\partial_{t}\psi-\sigma^{m}\mathcal{D}_{m}\psi+\mu\,\sigma_{2}\psi^{*}=0\label{eq:weyleq2}\,.
\end{equation}
Here the matrices $\sigma^{m}$ are not constant, but given by
\begin{equation}
\sigma^{m}=e_{a}^{m}\sigma^{a}\,,
\end{equation}
where $\sigma^{a}$ are the pauli matrices \eqref{paulimatrices}, and $e_{a}^{m}$ denote
the inverse of the dreibeins on $S^{3}$, which can be defined in
terms of the right invariant Maurer-Cartan forms on $SU(2)$ \eqref{dualforms} as
\begin{equation}\label{dreibeins3}
e^{a}=\frac{L}{2}\omega_{R}^{a}\;,\quad\delta_{ab}e_{m}^{a}e_{n}^{b}=L^2 h_{mn}\,.
\end{equation}
The covariant derivative on the sphere $\mathcal{D}_{m}$ acts on
a spinor in the form
\begin{equation}\label{spincon}
\mathcal{D}_{m}\psi=\left(\partial_{m}+\Gamma_{m}\right)\psi\;,\quad\Gamma_{m}=\frac{1}{8}\left[\sigma_{a},\sigma_{b}\right]e_{n}^{a}\nabla_{m}e^{nb}\,,
\end{equation}
where $\nabla_{m}$ is the covariant derivative with respect
to the affine connection on $S^{3}$. 

In terms of \eqref{majorana}, Eq. \eqref{eq:weyleq2} can be naturally rewritten as a Dirac equation in chiral representation 
\begin{equation} \label{eq:diraceq}
\left(\begin{array}{cc}
-\mu & i\left(\partial_{t}+\sigma^{m}\mathcal{D}_{m}\right)\\
i\left(\partial_{t}-\sigma^{m}\mathcal{D}_{m}\right) & -\mu
\end{array}\right)\left(\begin{array}{c}
\psi\\
-i\sigma^{2}\psi^{*}
\end{array}\right)=0\,.
\end{equation}
Then, the Klein-Gordon equation is obtained by applying the
operator 
\begin{equation}
\left(\begin{array}{cc}
\mu & i\left( \partial_{t}+\sigma^{n}\mathcal{D}_{n}\right)\\
i\left(\partial_{t}-\sigma^{n}\mathcal{D}_{n}\right) & \mu
\end{array}\right)
\end{equation}
on \eqref{eq:diraceq}. Using \eqref{paulirel}, the vielbein potulate $\mathcal{D}e^{a}=0$, and the fact that on
a vector-spinor $\chi_{m}$ the covariant derivative $\mathcal{D}_m$ acts like $\mathcal{D}_{m}\chi_{n}=\nabla_{m}\chi_{n}+\Gamma_{m}\chi_{n}$, this yields
\begin{equation}\label{KGweyl}
\left[-\partial_{t}^{2}+\frac{1}{L^2}h^{mn}\left(\nabla_{m}+\Gamma_{m}\right)\left(\partial_{n}+\Gamma_{n}\right)+\frac{R_{S^{3}}}{4}-\mu^{2}\right]\psi=0\,,
\end{equation}
where $R_{S^{3}}=6/L^{2}$ is the Ricci scalar of the three-sphere in \eqref{metricrxs3}.

\providecommand{\href}[2]{#2}\begingroup\raggedright\endgroup

\end{document}